\documentclass[twocolumn]{pasj02}

\usepackage{lineno}

\Received{2025 Jan 22}
\Accepted{2025 Feb 25}
\Published{?}

\usepackage{natbib}
\usepackage{graphicx}	
\usepackage{color}
\usepackage{tabularx}
\usepackage{dcolumn}
\usepackage{booktabs}
\usepackage{soul}

\newcommand{\sun}{\odot}
\newcommand{\rmxaa}{RMxAA}

\newcommand{\oi}{[O\,{\sc i}]}
\newcommand{\oii}{[O\,{\sc ii}]}
\newcommand{\oiii}{[O\,{\sc iii}]}

\newcommand{\NI}{[N\,{\sc i}]}
\newcommand{\nii}{[N\,{\sc ii}]}
\newcommand{\niii}{[N\,{\sc iii}]}

\newcommand{\sii}{[S\,{\sc ii}]}
\newcommand{\siii}{[S\,{\sc iii}]}
\newcommand{\siv}{[S\,{\sc iv}]}

\newcommand{\hei}{He\,{\sc i}}
\newcommand{\heii}{He\,{\sc ii}}

\newcommand{\neii}{[Ne\,{\sc ii}]}
\newcommand{\neiii}{[Ne\,{\sc iii}]}

\newcommand{\ariii}{[Ar\,{\sc iii}]}
\newcommand{\ariv}{[Ar\,{\sc iv}]}

\newcommand{\cliii}{[Cl\,{\sc iii}]}

\newcommand{\feii}{[Fe\,{\sc ii}]}
\newcommand{\arii}{[Ar\,{\sc ii}]}
\newcommand{\sIiii}{[Si\,{\sc iii}]}

\newcommand{\feiii}{[Fe\,{\sc iii}]}

\newcommand{\ciii}{C\,{\sc iii}}

\newcommand{\ha}{H$\alpha$}
\newcommand{\hb}{H$\beta$}

\newcommand{\hi}{H\,{\sc i}}

\newcommand{\kms}{km\,s$^{-1}$}
\newcommand{\te}{$T_{\rm e}$}
\newcommand{\Ne}{$n_{\rm e}$}
\newcommand{\ergs}{erg\,s$^{-1}$\,cm$^{-2}$\,{\AA}$^{-1}$}
\newcommand{\ergsF}{erg\,s$^{-1}$\,cm$^{-2}$}
\newcommand{\ergsm}{erg\,s$^{-1}$\,cm$^{-2}$\,{\micron}$^{-1}$}

\begin{document} 

\title{ 
Seimei KOOLS-IFU mapping of the gas and dust distributions in Galactic PNe: the origin and evolution of DdDm\,1}

\author{Masaaki \textsc{Otsuka}\altaffilmark{1}\altemailmark
}
\altaffiltext{1}{Okayama Observatory, Kyoto University, Honjo, Kamogata, Asakuchi, Okayama 719-0232, Japan}
\email{otsuka@kusastro.kyoto-u.ac.jp}
\orcid{0000-0001-7076-0310}


\maketitle

\begin{abstract}

We conduct a detailed study of the planetary nebula (PN) DdDm\,1 in the Galactic halo field. 
DdDm\,1 is a metal-deficient and the most carbon-poor PN (${\rm C/O} =0.11 \pm 0.02$) identified in the Galaxy.
We aim to verify whether it evolved into a PN without experiencing the third dredge-up (TDU) during the thermal pulse asymptotic giant branch (AGB) phase and 
to investigate its origin and evolution through accurate measurements of the physical parameters of the nebula and its central star.
We perform a comprehensive investigation of DdDm\,1 using multiwavelength spectra. 
The KOOLS-IFU emission line images achieve $\sim0.9$\,arcsec resolution, resolving the elliptical nebula and revealing a compact spatial distribution of the {\feiii} 
line compared to the {\oiii} line, despite their similar volume emissivities.
This indicates that iron, with its higher condensation temperature than oxygen, is easily incorporated into dust grains such as silicate, 
making the iron abundance estimate prone to underestimation.
Using a fully data-driven approach, we directly derive ten elemental abundances, the gas-to-dust mass ratio, and the gas and dust masses based on our own 
heliocentric distance scale (19.4\,kpc)
and the emitting volumes of gas and dust. 
Our analysis reveals that DdDm\,1 is a unique PN evolved from a single star with an initial mass of $\sim1.0$\,M$_{\sun}$ and a metallicity $Z$ of 0.18\,Z$_{\sun}$. 
Thus, DdDm\,1 is the only known PN that is confirmed to have evolved without experiencing TDUs. 
The photoionization model reproduces all observed quantities in excellent agreement with predictions from AGB nucleosynthesis, post-AGB evolution, and AGB dust production models. 
Our study provides new insights into the internal evolution of low-mass and metal-deficient stars like DdDm\,1 and 
highlights the role of PN progenitors in the chemical enrichment of the Galaxy.
\end{abstract}


\section{Introduction}
\label{S-intro}

Planetary nebulae (PNe) represent the terminal evolutionary stage of low- to intermediate-mass stars with initial masses ranging 
from $\sim$1\,M$_{\sun}$ to $\sim$8\,M$_{\sun}$ \citep[e.g.,][]{2000oepn.book.....K}. 
These objects provide invaluable insights into stellar evolution, nucleosynthesis, mass-loss history, and even the chemical enrichment of galaxies. 

In the Milky Way, over 3000 PNe have been cataloged, the majority belonging to the Galactic disk \citep{2016JPhCS.728c2008P}. 
In contrast, a smaller subset of 13 PNe in the Galactic halo has been identified \citep{2023PASJ...75.1280O}.
Halo PNe are notable for their low metallicity, with an average [Ar/H] of $\sim-1.33$ \citep[see table\,1 of][]{2023PASJ...75.1280O}. 
This characteristic, along with the retention of asymptotic giant branch (AGB) nucleosynthesis signatures, makes halo PNe invaluable for studying the early Galactic environment. 
Studies of objects such as 
K\,648 \citep{1979RMxAA...4..341T,1992A&A...265..757P,2002A&A...381.1007R,2015ApJS..217...22O,2017ApJ...836...93J}, 
H\,4-1 \citep{1979RMxAA...4..341T,2015ApJS..217...22O,2023PASJ...75.1280O}, 
BoBn\,1 \citep{1991PASP..103..865P,1993RMxAA..27..175P,2006MNRAS.369..875Z,2008ApJ...682L.105O,2010ApJ...723..658O}, 
and DdDm\,1 \citep{1987MNRAS.224..761C,1992A&A...265..757P,2005MNRAS.362..424W,2008ApJ...680.1162H,2009ApJ...705..509O} 
have significantly advanced our understanding of their unique evolutionary pathways and low metallicity, offering rare insights into the conditions of the early Galaxy.

DdDm\,1 is one of the field halo PNe, with its metallicity close to the average [Ar/H] value among halo PNe. 
The basic physical parameters of DdDm\,1 are summarized in table\,1 of \citet{2009ApJ...705..509O}. 
Unlike most other halo PNe, which are predominantly C-rich, DdDm\,1 is the only known O-rich (i.e., C/O $<1$) halo PN and exhibits the most C-poor abundance among the Galactic PNe. 
The C-poorness and normal N and O abundances suggest that DdDm\,1 experienced neither efficient third dredge-up (TDU) events nor hot bottom burning during the thermal pulse AGB phase.
Consequently, the studies by \citet{1987MNRAS.224..761C}, \citet{2008ApJ...680.1162H}, and \citet{2009ApJ...705..509O} have concluded that DdDm\,1 evolved from a progenitor star with 
an initial mass of $\sim1$\,M$_{\sun}$.

However, this conclusion lacks sufficient theoretical validation. 
At that time, neither theoretical AGB nucleosynthesis models nor post-AGB evolution models were available for stars with an initial mass of $\sim1$\,M$_{\sun}$ and a metallicity $Z$ of $\sim0.1$\,Z$_{\sun}$.
As a result, we could not rigorously compare the observed elemental abundances of DdDm\,1 with the values predicted by such models.
Additionally, there has been insufficient investigation into whether the central star's effective temperature, luminosity, surface gravity, and the amount of gas/dust masses ejected during its evolution, 
as estimated from photoionization models, align with the predictions of post-AGB evolutionary models for such stars. 
Furthermore, these nebular and central star's parameters and the distance to DdDm\,1 derived from photoionization models are 
based on physical parameters derived from nebular spectra obtained through limited wavelength coverage and slit positions. 
There is no guarantee that the physical parameters derived from such data truly reflect the intrinsic properties of DdDm\,1. 
Therefore, the origin and evolutionary history of DdDm\,1 remain unclear.

Over the past decade, significant advancements have been made in AGB nucleosynthesis and post-AGB stellar evolutionary models as well as in observational instruments such 
as integral field spectrograph like Seimei/KOOLS-IFU. 
These instruments can capture wide wavelength spectra of entire nebulae in a single exposure. 
Leveraging these developments, we revisit the origin and evolution of DdDm\,1.

The primary aim of this paper is twofold. 
First, we verify whether DdDm\,1 experienced TDUs by comparing observed elemental abundances 
with theoretical AGB model prediction. 
If proven, it would represent a unique and rare PN without experiencing TDUs. 
This finding would raise questions about the evolution and particularly carbon production in PN progenitors. 
Second, we seek to create a coherent and detailed picture of DdDm\,1. 
We construct a comprehensive photoionization model that achieves perfect consistency between the observed quantities and the theoretical model predictions, based on a multiwavelength dataset.
Our model will allow us to derive the current status of the central star and the ejected gas/dust masses, shedding light on the unique properties of DdDm\,1.
Beyond this specific target, our study will offer broader implications for understanding the evolution of metal-poor stars and their contributions to the early Galaxy.

This paper is organized as follows. 
In sections \ref{S-data} and \ref{S-image}, we describe the Seimei/KOOLS-IFU observations, the other used dataset, and the data reduction processes. 
In section \ref{S-ext}, we perform detailed plasma diagnostics. 
A comprehensive abundance analysis is presented in section \ref{S:abund}, 
followed by empirical calculations of gas and dust masses and the gas-to-dust mass ratio in section \ref{S-gdr}. 
In section \ref{S-dis}, we discuss the evolutionary history of DdDm\,1.  
Lastly, we summarize the present work in section \ref{S-sum}.

\begin{figure}
\begin{center}
\includegraphics[width=\columnwidth]{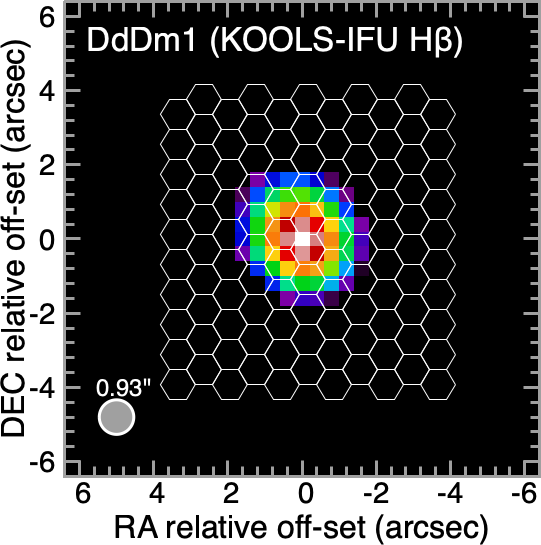}
\end{center}
\caption{
A dither position of the packed fiber array in KOOLS-IFU observations is overlaid on the PSF-deconvolved {\hb} emission 
image derived from the KOOLS-IFU data cube. The intensity scale is shown in logarithmic units, ranging from minimum to maximum. 
The gray circle represents the spatial resolution of the image. Each hexagon denotes the position and dimensions of the 110 fibers.
{Alt text: A PSF-deconvolved {\hb} emission image from the KOOLS-IFU dataset, overlaid with dithered positions of the packed fiber array.}
\label{F:koolsfv}} 
\end{figure}

\section{Dataset and reduction}
\label{S-data}

We explain the dataset and data reduction processes below.  
The logs of the spectroscopic observations are summarized in table\,\ref{T-obslog}.

\subsection{Seimei/KOOLS-IFU observation} 
\label{S:kools}

\begin{table}
\caption{Summary of spectroscopic observations for DdDm\,1. \label{T-obslog}}
\centering
\begin{tabularx}{\columnwidth}{@{\extracolsep{\fill}}l@{\hspace{5pt}}c@{\hspace{3pt}}D{-}{-}{-1}c@{\hspace{3pt}}c}
\midrule	 
Date & Telescope/Inst. & \multicolumn{1}{c}{Range ({\micron})} &$R$\\
\midrule
2024/07/21    & Seimei/KOOLS-IFU   &0.400~-~0.853      &$\sim820$     \\
2024/07/21    & Seimei/KOOLS-IFU   &0.560~-~1.010      &$\sim900$     \\
1990/09/06    & \emph{IUE}/SWP     &0.115~-~0.198      &$\sim300$   \\
1984/09/01,04 & \emph{IUE}/SWP     &0.115~-~0.198      &$\sim300$   \\
2008/07/08 &Subaru/HDS &0.39~-~0.66 &$\sim72000$\\
2008/07/08 &Subaru/HDS &0.30~-~0.46 &$\sim72000$\\
2006/05/11 & \emph{Spitzer}/IRS SL &5.2~-~14.5         &$\sim100$    \\
2006/05/11 & \emph{Spitzer}/IRS SH, LH &9.9~-~37.2         &$\sim600$   \\
\midrule
\end{tabularx}
\end{table}

We performed the dither-mapping observation (figure\,\ref{F:koolsfv}) using the 
Kyoto Okayama Optical Low-dispersion Spectrograph with optical-fiber integral field unit 
\citep[KOOLS-IFU;][]{2019PASJ...71..102M} attached to one of the two Nasmyth foci of 
the Kyoto University Seimei 3.8\,m telescope \citep{2020PASJ...72...48K}.

The observation was carried out under clear and stable sky conditions on 2024 July 21. 
The seeing, measured by the Shack–Hartmann camera, was $\sim1{\arcsec}$ in the $R$-band at full width at half maximum (FWHM). 
The data were obtained at an airmass range of $1.01-1.09$ (average 1.04) for the VPH-blue setting and $1.19-1.51$ (average 1.35) for the VPH-red setting. 
We performed $7\times300$\,sec and $3\times360$\,sec exposures for the blue observations, 
and $6\times300$\,sec and $3\times360$\,sec exposures for the red observations, respectively. 
In each exposure, the data were taken with a sub-fiber pitch shift, increasing the spatial sampling from 
0.8{\arcsec}\,spaxel$^{-1}$ to 0.4{\arcsec}\,spaxel$^{-1}$. 
Additionally, off-source frames were taken for background sky subtraction.

To perform flux calibration and remove telluric absorption,  
we observed HD192281 (O4.5V(n)) every 30\,min using the same dithering pattern as employed for DdDm\,1.  
To further evaluate light loss in the KOOLS-IFU field, BD+39{\degree}\,3026 (F5) nearby DdDm\,1 was observed at every increase of 0.1 in airmass.  
We also obtained instrumental flat, twilight sky, and Hg/Ne/Xe-arc frames for baseline calibration.

\subsection{Seimei/KOOLS-IFU data reduction}

We reduced the data using our codes and {\sc IRAF} \citep{1986SPIE..627..733T} by following \citet{2022MNRAS.511.4774O}.  
Wavelength calibration was performed using over 40 Hg/Ne/Xe-arc lines that cover the full range of effective wavelengths.  
The resultant RMS errors in wavelength determination were 0.09\,{\AA} for the VPH-blue and 0.10\,{\AA} for the VPH-red. We adopted a constant plate scale of 2.5\,{\AA}. 
The average spectral resolution ($R$, defined as $\lambda$/Gaussian FWHM) measured from the arc lines was $\sim820$ and $\sim900$  
for the VPH-blue and VPH-red spectra, respectively. We calibrated the extinction caused by airmass using the extinction magnitude-to-airmass ratio curve measured at Okayama Observatory in 2022.  
This curve was derived over the wavelength range of $4000-10000$\,{\AA}, with measurements taken at intervals of 2.5\,{\AA}.

We obtained an RA-DEC-wavelength datacube using our codes for image-data reconstruction based  
on the drizzle technique. We corrected the spatial displacements caused by atmospheric dispersion  
at each wavelength by following \citet{2022MNRAS.511.4774O}.  
We adopted a constant plate scale of 0.40{\arcsec} per spaxel, which provides a good signal-to-noise ratio (SNR)  
in each spaxel. During this procedure, we also generated the SNR map required for uncertainty evaluation  
for each wavelength image in the subsequent analyses.  
We evaluated the off-center value in RA and DEC relative to a reference position  
by measuring the center coordinates of the standard star in the 3-D datacube:  
the offset did not exceed 0.02{\arcsec} (0.05\,pixel) in both directions over the wavelengths.

For flux calibration and telluric removal, we synthesized the non-LTE theoretical spectrum of  
HD192281 by fitting its archived Keck-II/ESI spectrum \citep[$3850 - 9360$\,{\AA}, $R\sim6200$ at 5020\,{\AA};][]{2002PASP..114..851S} 
with the {\sc Tlusty} code \citep{1988CoPhC..52..103H}.  
We reduced the raw data (Program ID: W171, PI: Ferre-Mateu) downloaded from the Keck Observatory Archive webpage.  
By following a traditional fitting method with the aid of a genetic algorithm, we successfully determined seven 
elemental abundances\footnote{We derived $12 + \log_{10}$\,$n$(X)/$n$(H) = 11.30, 8.12, 8.30, 7.91, 7.06, 5.63, 5.68, and 6.95\,dex in He/C/N/O/Mg/Al/Si/Fe.}, 
surface gravity ($\log_{10}\,g) = 3.53$\,cm\,s$^{-2}$, effective temperature 
($T_{\rm eff}) = 38051$\,K, and rotational velocity ($v\sin\,i$) = 298\,{\kms}.  
We calculated the color excess between {\it B} and {\it V} bands, $E(B-V) = 0.628$, toward HD192281  
with a total-to-selective extinction $R_{\rm V} = 3.1$ by comparing the synthesized spectrum and the  
{\it Gaia} BP/RP spectrum 
after spectral resolution matching. In this procedure, we adopted  
the reddening law of \citet{1989ApJ...345..245C}. Flux calibration and telluric removal were simultaneously performed using the function  
derived by comparing this reddened {\it telluric-free} theoretical spectrum and the 1-D spectrum extracted from the observed  
3-D datacube.

\subsection{Seimei/KOOLS-IFU data point-spread function}

We evaluated the achieved spatial resolution by measuring the Moffat function’s FWHM of the standard stars' PSF at every wavelength.  
The Moffat function is represented as follows:  

\begin{eqnarray}
I(r) &=& \left[1 + \left(r/{\alpha}\right)^{2}\right]^{-\beta}, \\
{\alpha} &=& {\rm FWHM}\left[2^{\left(1/{\beta}\right)} - 1 \right]^{-0.5},  
\label{E:moffat}
\end{eqnarray}

\noindent where $I(r)$ is the radial profile of the PSF of the standard stars,  
and $\beta$ is the shape parameter and is fixed to 4.4 at all wavelengths.  
To correct for image quality degradation due to airmass, we adopted the equation established  
by \citet{2022MNRAS.511.4774O}. Consequently, the expected Moffat PSF's FWHM of DdDm\,1's KOOLS-IFU images  
in arcsec can be expressed as a function of $\lambda$ in {\micron} using equation\,(\ref{E:psf}).  

\begin{eqnarray}
{\rm FWHM} = \left\{ \begin{array}{ll}
(1.67\pm0.01)\,(\lambda/0.55)^{-0.18\pm0.01} & \mbox{(blue)}\\
(1.90\pm0.01)\,(\lambda/0.55)^{-0.17\pm0.01} & \mbox{(red)}
\end{array} \right.
\label{E:psf}
\end{eqnarray}

After performing PSF matching in all spectral bins in  
the datacube using equation\,(\ref{E:psf}), we carried out spectral analysis.

\subsection{Subaru/HDS spectrum of the central star 
\label{S-hdsspec}} 

We investigate the natures of the central star with its high-dispersion echelle spectra taken using the High Dispersion Spectrograph \citep[HDS;][]{2002PASJ...54..855N} attached to 
one of Nasmyth foci of the Subaru 8.2\,m telescope atop of Mt. Mauna Kea in Hawaii. 
We downloaded the archival raw data (PI: A.~Tajitsu) from the Subaru Mitaka Okayama Kiso Archive (SMOKA). 
The observation was conducted under stable sky condition with a seeing of $\sim0.6${\arcsec} measured from the HDS slit-viewer camera. 
The slit-entrance has dimensions of 0.5{\arcsec} in width and 3.0{\arcsec} in length. 
Spectra were obtained using the blue and red cross dispersers, covering wavelength ranges of $2974-4610$\,{\AA} and $3945-5210$\,{\AA}, respectively. 
The image-de-rotators to keep position angle (P.A.) of 0{\degree} were used for both settings. 
For the red setting, we use an atmospheric dispersion corrector (ADC). 
We took a single 1800\,sec exposure for both settings. 
We reduced the data with our codes and {\sc IRAF} packages. 
We adopted a constant plate scale of 0.025\,{\AA}\,pixel$^{-1}$.

\subsection{UV \emph{IUE} and mid-infrared \emph{Spitzer} spectra \label{S-irspec}} 

We downloaded the SWP23840, 23872 (IUE program ID: GM115, PI: R.E.S.\,Clegg), and SWP39590 (ID: PNMMP, PI: M.\,P\~{e}na) 
obtained by the International Ultraviolet Explorer (\emph{IUE}) from the Mikulski Archive for Space Telescopes (MAST). 
The \emph{IUE} spectrum was used to derive the C$^{2+}$ and N$^{2+}$ ionic abundances from their collisionally excited lines (CELs). The size of the slit (9.9{\arcsec}~$\times$~22{\arcsec}) covered the entire nebula of DdDm\,1. 
We combined these spectra into a single $1150-1980$\,{\AA} spectrum using an SNR-weighted averaging method. 
The spectrum was then scaled to match the photometry measured from the Galaxy Evolution Explorer
\citep[\emph{GALEX};][]{2017ApJS..230...24B} FUV channel image ($F_{\rm FUV} = (9.90\pm0.10)\times10^{-14}$\,{\ergs}).

We also used the archived mid-IR {\it Spitzer}/Infrared Spectrograph \citep[IRS;][]{2004ApJS..154...18H} short-low (SL), short-high (SH), and long-high (LH) spectra (AOR key: 16808448 and 16808704, PI: K.~B.~Kwitter).  
The size of the slit was $3.7\arcsec \times 57\arcsec$ in SL, $4.7\arcsec \times 11.3\arcsec$ in SH, and $11.1\arcsec \times 22.3\arcsec$ in LH.  These spectra were used to derive the Ne$^{+,2+}$, S$^{2+,3+}$, Ar$^{+,2+}$, and Fe$^{2+}$ ionic abundances from their fine-structure lines and to examine dust features. 
We degraded the spectral resolution of the SH and LH spectra to match that of the SL spectrum and then combined  
these three spectra into a single spectrum. Finally, we scaled the flux density to match the mid-IR  
\emph{AKARI}/IRC 18\,{\micron} band flux density \citep[$F_{18\,{\micron}} = (2.16\pm0.14)\times10^{-12}$\,{\ergsm}
;][]{2010A&A...514A...1I}.

\begin{figure*}
\includegraphics[width=\textwidth]{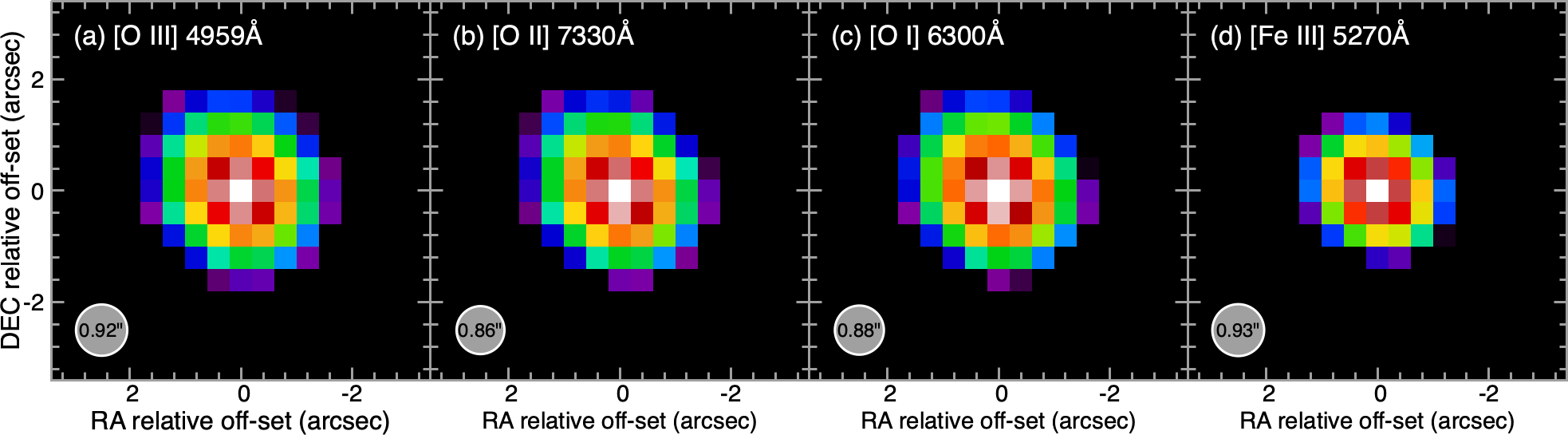}
\caption{
Selected PSF-deconvolved KOOLS-IFU emission-line images. 
The intensity is presented on a logarithmic scale, spanning from the minimum to the maximum values. 
The achieved spatial resolution is indicated by the gray circle. 
The spatial distribution of the {\feiii}\,5270\,{\AA} emission is much smaller in radius 
compared to that of the {\oiii}\,4959\,{\AA} emission, suggesting that a significant fraction of iron is locked in dust grains.
See the main text for further details.
{Alt text: The PSF-deconvolved images of the four emission lines.
 The spatial distribution of the {\feiii} line is more compact than that of {\oiii}, suggesting that a significant fraction of iron is locked in dust grains.
}
\label{F:map-kools}} 
\end{figure*}

\section{Results of the KOOLS-IFU observations}
\label{S-image}

Before starting analyses, we check the accuracy of the data. 
We compare the flux density directly measured from the Pan-STARRS 
$r$-band image (downloaded from the Pan-STARRS1 data archive) with that derived from the synthesized $r$-band image 
created from the KOOLS-IFU datacube using the Pan-STARRS $r$-band filter transmission curve. The difference between 
the measured flux densities from both the two sources is only $\sim0.7$\,$\%$, with values of $(7.74 \pm 0.02) \times 10^{-15}$\,{\ergs} for Pan-STARRS 
and $(7.69 \pm 0.01) \times 10^{-15}$\,{\ergs} for KOOLS-IFU. Therefore, we can confidently conclude 
that our KOOLS-IFU data achieves sufficient accuracy in photometry.

\subsection{Spatial distribution of emission lines \label{S:maps}}

In figure\,\ref{F:map-kools}, we present the images of the {\oiii}\,4959\,{\AA}, {\oii}\,7330\,{\AA}, {\oi}\,6300\,{\AA}, and 
{\feiii}\,5270\,{\AA} lines, which are used to examine the ionization stratification and the spatial deficiency of ionized iron.
These images were created by measuring the line fluxes in each spectrum for every spaxel using the automated line-fitting algorithm
{\sc Alfa} \citep{Wesson:2016aa}, followed by a Lucy-Richardson PSF-deconvolution with fifteen iterations.  
To our knowledge, such emission-line images have never been presented before.

DdDm\,1 is apparently very compact and exhibits a slightly elongated nebula with a position angle of $47{\degree}-52{\degree}$. 
This is consistent with the findings of \citet{2009ApJ...705..509O} and \citet{2008ApJ...680.1162H}, who demonstrated the elliptical nebular morphology of DdDm\,1 using the broadband \emph{HST}/WFPC1 F675W image. 
The three-sigma Gaussian radius along the major axis is estimated to be 1.03{\arcsec}, 1.10{\arcsec}, and 1.18{\arcsec}  
in {\oiii}, {\oii}, and {\oi}, respectively. 
These values correspond to $3 \times (\sigma_{\rm obs}^{2} - \sigma_{\rm res}^2)^{0.5}$, where $\sigma_{\rm obs}$ denotes the Gaussian sigma of the emission-line distribution after PSF deconvolution, and $\sigma_{\rm res}$ represents the Gaussian sigma of the achieved spatial resolution. 
This calculation evaluates the intrinsic spatial extent of the emission-line distribution by accounting for and removing the broadening effects caused by seeing and instrumental resolution. 
Using these sub-arcsecond-resolution images, we confirm the ionization stratification, i.e., the nebular radius decreases as  the ionization degree increases.

It is well known that the deficiency of elemental iron abundance relative to the Sun is  
consistently significantly lower than that of other elements, such as oxygen.  
This has been confirmed in ionized nebulae including PNe.  
Indeed, the present work derived [O/H] = $-0.64 \pm 0.06$ and [Fe/H] = $-1.32 \pm 0.13$ from our abundance analysis (subsection\,\ref{S:eabund1}).  
This is because iron is a refractory element, while oxygen is a volatile element.  
Since iron atoms tend to exist as dust grains, we expect that the spatial distributions of ionized atomic iron emission lines in ionized nebulae are more compact than those of volatile elements, even with the same ionization degrees.  
We verify this through measurements of the spatial distributions of {\oiii}\,4959\,{\AA} (figure\,\ref{F:map-kools}a) and the {\feiii}\,5270\,{\AA} (figure\,\ref{F:map-kools}d). 
The spatial distribution of the {\feiii}\,5270\,{\AA} emission is noticeably more compact than that of the {\oiii}\,4959\,{\AA} emission; 
the three-$\sigma$ Gaussian radius along the major axis is 0.61{\arcsec} for {\feiii}\,5270\,{\AA}, which is approximately half of the 1.03{\arcsec} radius measured for {\oiii}\,4959\,{\AA}. At the {\oiii} electron temperature {\te}({\oiii}) and 
the {\cliii} electron density {\Ne}({\cliii}) (subsection\,\ref{S:tene}), the volume emissivity of the {\feiii}\,5270\,{\AA} line is $(2.29 \pm 0.81) \times 10^{-21}$\,erg\,s$^{-1}$\,cm$^{3}$, while 
that of the {\oiii}\,4959\,{\AA} line is $(1.86 \pm 0.03) \times 10^{-21}$\,erg\,s$^{-1}$\,cm$^{3}$. 
Therefore, we infer that their spatial extent is not attributable to the difference in volume emissivity but rather to the difference in the spatial distributions of Fe$^{2+}$ and O$^{2+}$. 
This supports our expectation that iron emission lines are spatially more compact due to dust formation.

In many studies, the Fe abundance in ionized gas is determined using the Fe$^{2+}$ abundance and the O/O$^{2+}$ ratio. 
However, the spatial distribution differences between Fe$^{2+}$ and O$^{2+}$ must be considered; 
otherwise, the Fe abundance may inevitably be underestimated. 
Careful attention is required when interpreting the Fe abundance estimates.

\subsection{Spatially-integrated 1D-spectrum}

\begin{figure*}
\includegraphics[width=\textwidth]{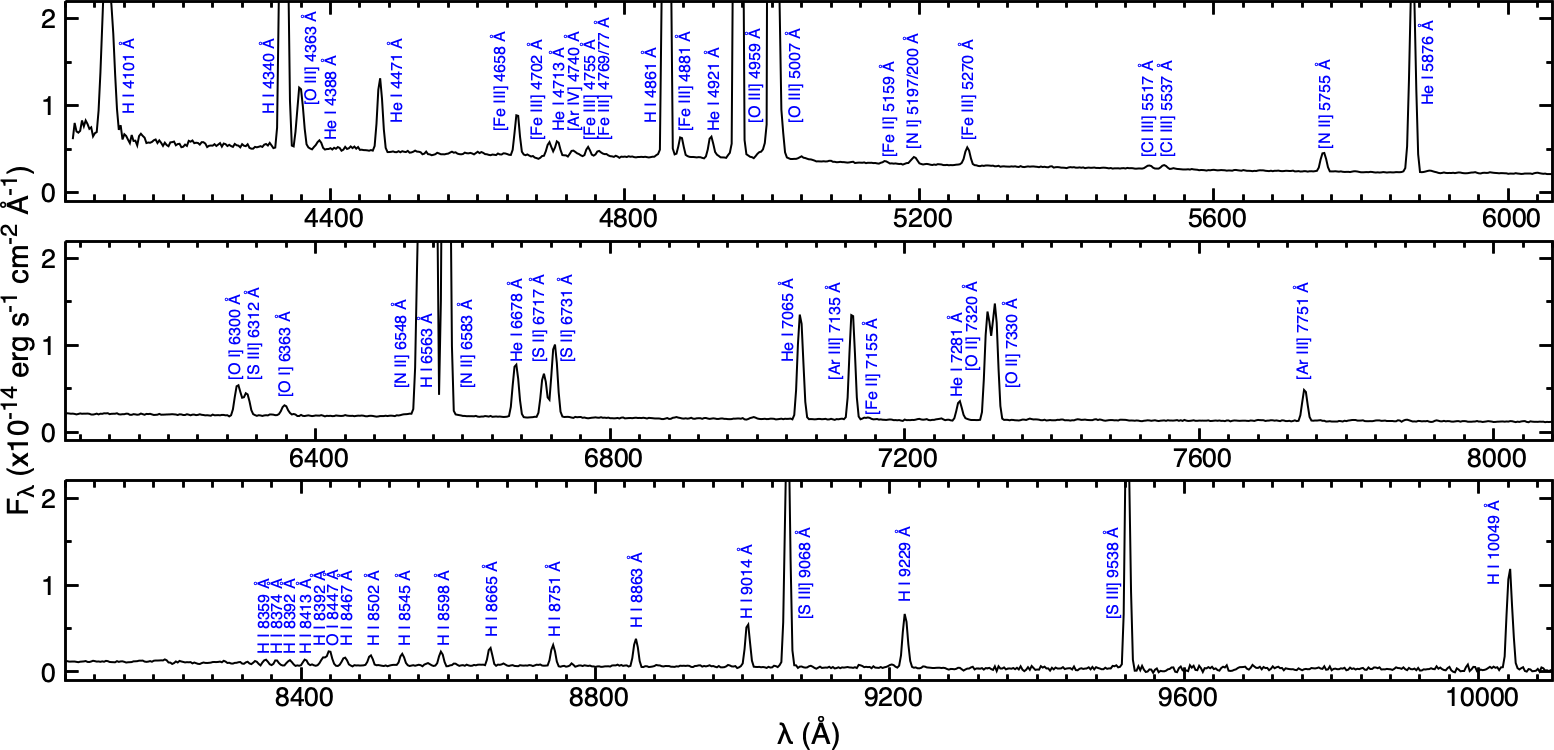}
\caption{The KOOLS-IFU 1-D spectrum. Prominent lines among the identified emission lines are indicated in blue.  
The wavelengths are given in air and include a heliocentric correction. 
The dark-gray shaded region indicates the one-sigma uncertainty at each wavelength. 
{Alt text: A spatially integrated 1-D spectrum generated from the KOOLS-IFU datacube. The horizontal and vertical axes are wavelength and flux density, respectively.}
\label{F:spec-kools}} 
\end{figure*}

As shown in figure\,\ref{F:spec-kools}, we analyze the spatially integrated 1-D spectrum obtained from the KOOLS-IFU datacube. 
The spectrum is presented in air wavelengths with heliocentric corrections applied, and the heliocentric radial velocity is 
measured as $-308.2 \pm 8.2$\,{\kms}, consistent with \citet[$-309.14 \pm 1.22$\,{\kms}]{2009ApJ...705..509O}. 
Numerous prominent emission lines are clearly detected, while no carbon recombination lines are observed. 
This spectrum serves as the basis for the detailed analyses presented in the following sections.

\section{Mid-infrared features}

\begin{figure}
\includegraphics[width=\columnwidth]{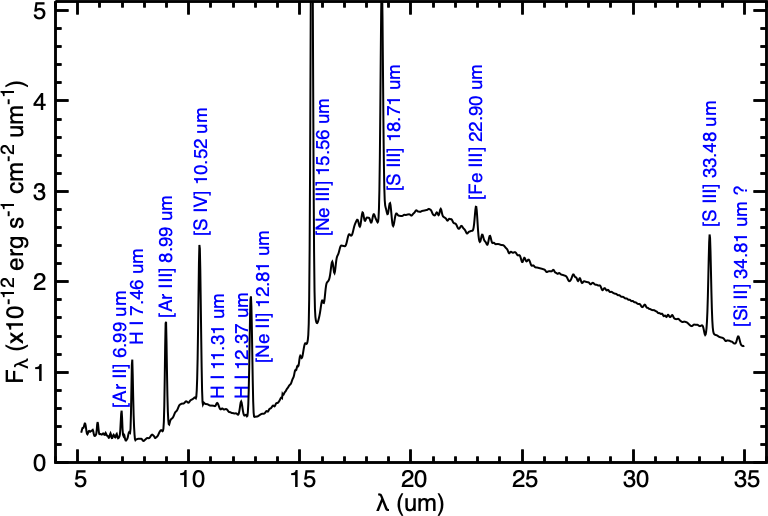}
\caption{The \emph{Spitzer}/IRS spectrum. 
The detected emission lines are indicated in blue. The wavelengths are given in vacuum. 
{Alt text: The \emph{Spitzer}/IRS spectrum. 
The horizontal and vertical axes are wavelength and flux density, respectively.
}
\label{F:spec-spitzer}} 
\end{figure}

The \emph{Spitzer}/IRS spectrum (figure\,\ref{F:spec-spitzer}) clearly shows two broad features  attributed to amorphous silicate grains; 
the features centered at 9\,{\micron} and 18\,{\micron} are due to the Si-O stretching mode and the O-Si-O bending mode, respectively.
Crystalline silicate emission are not detected in the spectrum around 11, 19.5, 23--24, 28, and 33.6\,{\micron}.
Crystallization requires a re-heating process of amorphous dust. 
In the case of DdDm\,1, this process is likely insufficient, resulting in the dominance of amorphous silicates.
No carbon-based dust grains and molecules are not confirmed in the IRS spectrum. 
Therefore, DdDm\,1 is certainly a O-rich dust PN.

\section{Dust extinction and line flux measurement\label{S-ext}}

We measure the line fluxes of atomic species in the 1-D KOOLS-IFU, \emph{IUE}, and \emph{Spitzer} spectra by fitting multiple Gaussian functions to their line profiles. 
The observed flux $F$($\lambda$) (and flux density $F_{\lambda}$) is reddened by both the interstellar and also the circumstellar dust. 
We obtain the dust extinction-free flux $I$($\lambda$) (and flux density $I_{\lambda}$) using equation\,(\ref{eq-1}):

\begin{eqnarray}
  I(\lambda) &=& F(\lambda)~\cdot~10^{c({\rm H\beta})(1 + f(\lambda))},
   \label{eq-1}
\end{eqnarray}

\noindent where $c$({\hb}) ($= \log_{10}$\,$I$({\hb})/$F$({\hb})) is the line-of-sight logarithmic reddening coefficient at {\hb}\,4861.33\,{\AA}.
The $c$({\hb}) value can be expressed as $1.45\, E(B-V)$ under $R_{\rm V} = 3.1$ and the reddening law of \citet{1989ApJ...345..245C}.  
The $E(B-V)$ value within a $2{\degree}\times2{\degree}$ area centered on the coordinates of DdDm\,1 is $(9.8 \pm 0.7) \times 10^{-3}$,  
according to the Galaxy extinction map of \citet{2011ApJ...737..103S}.  
Thus, the $c$({\hb}) value is calculated to be $0.014 \pm 0.001$.  
Note that this value originates from the average global interstellar dust over a large solid angle around DdDm\,1,  
as the Galaxy extinction map is not high enough to resolve individual objects.  
Therefore, it is necessary to calculate the $c$({\hb}) value stemming from the combined contributions of interstellar and circumstellar dust.
This can be achieved by comparing the observed extincted {\hi} line fluxes with the corresponding extinction-free {\hi} line fluxes,  
calculated using the {\hi} emissivity of \citet{Storey:1995aa} under the assumption of Case B.   
For this purpose, we employed {\hi} lines detected in the KOOLS-IFU spectrum.

The {\ha}/{\hb} ratio is often assumed to be a constant value (e.g., 2.85), and this assumption is commonly used to calculate the $c$({\hb}) value. 
While this assumption is widely adopted as it simplifies calculations, it is not always appropriate.
This is because the {\hi} line emissivity depends on electron temperature ({\te}) and electron density ({\Ne}), 
which cannot be determined without performing plasma diagnostics based on de-reddened line fluxes. 
Correcting line reddening based on incorrect assumptions and using the resulting physical quantities for further discussion is problematic and potentially misleading. 
To avoid such issues and to derive a self-consistent $c$({\hb}) value, {\te}, and {\Ne}, it is essential to calculate them iteratively within a calculation loop.    
By iteratively refining these parameters, this approach ensures that the derived $c$({\hb}) value aligns consistently with the {\hi} line emissivity determined 
by the plasma diagnostics, ultimately enabling more physically reliable interpretations.

This iterative loop for $c$({\hb}) derivation incorporates observed {\te} and {\Ne} diagnostic line fluxes, observed {\hi} line fluxes, and the theoretical {\hi} emissivity.  
For {\te} diagnostics, we use the {\oiii} lines at 4363, 4959, and 5007\,{\AA}.  
For {\Ne} diagnostics, the {\cliii} lines at 5517\,{\AA} and 5537\,{\AA} are selected, considering the ionization degree of {\oiii} and {\cliii}.  
We select the {\hi} lines at 4101, 4340, 4863, 6563, 8665, 8751, 8863, 9229, and 10049\,{\AA}, using their eight ratios relative to {\hb}.  
These {\hi} lines are chosen to optimize dust reddening corrections across the KOOLS-IFU wavelength range. 
We keep $R_{\rm V} = 3.1$ through our calculations.

The process begins by assuming an initial $c$({\hb}) value, set to 0 for the first iteration.  
Using this value, we apply reddening corrections to the {\oiii} and {\cliii} lines and determined {\te}({\oiii}) and {\Ne}({\cliii}).  
Theoretical {\hi} line ratios are then calculated based on the derived {\te}({\oiii}) and {\Ne}({\cliii}).  
These theoretical ratios are compared with observed values, yielding eight individual $c$({\hb}) estimates.  
Outliers are removed via three-sigma clipping, and the mean of the remaining $c$({\hb}) values is computed.  
From the mean $c$({\hb}) and the remaining values along with their one-sigma uncertainties, the reduced-$\chi^2$ value is calculated.  
This process is iteratively repeated, updating the $c$({\hb}) value until the reduced-$\chi^2$ reaches its minimum (2.43).

Through this iterative approach, we derive $c$({\hb}) = $0.136 \pm 0.013$, {\te}({\oiii}) = $11896 \pm 54$\,K, and {\Ne}({\cliii}) = $2536 \pm 759$\,cm$^{-3}$. 
The derived $c$({\hb}) value represents the combined contributions from both circumstellar and interstellar dust. 
Referring to  \citet{2019ApJ...887...93G}, we estimate $c$({\hb}) value solely from the interstellar dust to be $0.04\pm0.03$.
By subtracting this interstellar component from the total $c$({\hb}) value of $0.136 \pm 0.013$, we determine $c$({\hb}) = $0.098 \pm 0.041$ as the contribution from circumstellar dust. 
For the extinction correction in following sections, we adopt the total value of $c$({\hb}) = $0.136 \pm 0.013$.

In the 3-4th and 8-9th columns of table\,\ref{T-line}, we list the dereddened line fluxes normalized to the flux of {\hb}, which is set to an intensity of 100: 
the observed quantities $I$(obs) and ${\delta}I$(obs) represent the normalized dereddened line fluxes and their one-$\sigma$ uncertainty, 
respectively. The dereddened {\hb} line flux $I$({\hb}) for the entire nebula is measured to be ($2.54 \pm 0.08$)$\times10^{-12}$\,{\ergsF}.

\begin{table}
\caption{Derived {\te} and {\Ne} in DdDm\,1. \label{T-diagno}}
\centering
\begin{tabularx}{\columnwidth}{@{\extracolsep{\fill}}l@{\hspace{0pt}}c@{\hspace{0pt}}D{p}{\pm}{-1}
@{}
}
\midrule	 
Parameter &{\te} diagnostic line ratio&\multicolumn{1}{c}{Value (K)}\\
\midrule	 
{\te}({\oiii})   &$I$(4959\,{\AA}+5007\,{\AA})/$I$(4363\,{\AA})            &11896~p~54\\
{\te}({\siii})$_{\rm opt}$    &$I$(9068\,{\AA})/$I$(6312\,{\AA})  &13179~p~626\\
{\te}({\siii})$_{\rm opt/ir}$ &$I$(9068\,{\AA})/$I$(18.71\,{\micron}) &13129~p~810\\
{\te}({\ariii}) &$I$(7135\,{\AA}+7751\,{\AA})/$I$(8.99\,{\micron}) &11373~p~30\\
{\te}({\feiii}) &$I$(4881)/$I$(22.90\,{\micron}) &10911~p~1406\\
{\te}({\nii})  &$I$(6548\,{\AA}+6583\,{\AA})/$I$(5755\,{\AA})&13192~p~247\\\
{\te}({\hei})  &{\hei} lines&11793~p~1241\\
\midrule	 
Parameter &{\Ne} diagnostic line ratio&\multicolumn{1}{c}{Value (cm$^{-3}$)}\\
\midrule	 
{\Ne}({\siii})   &$I$(18.71\,{\micron})/$I$(33.48\,{\micron})&2656~p~722\\
{\Ne}({\cliii})  &$I$(5517\,{\AA})/$I$(5537\,{\AA})&2536~p~759\\
{\Ne}({\sii})    &$I$(6717\,{\AA})/$I$(6731\,{\AA})&3454~p~879\\

\midrule
\end{tabularx}
\end{table}

\section{Plasma diagnostics \label{S:abund}}

\subsection{{\te} and {\Ne} derivations \label{S:tene}}

Table\,\ref{T-diagno} summarizes the derived {\te} and {\Ne} values. 
The {\Ne} and {\te} values from CELs are determined at the intersection points of the diagnostic {\Ne}--{\te} curves for each ionization zone. 
The {\te}({\oiii}), {\te}({\ariii}), and {\te}({\feiii}) values correspond to the intersection points of their respective {\Ne}--{\te} curves with the one derived from {\Ne}({\cliii}). 
Even when the {\Ne}--{\te} curve derived from {\Ne}({\siii}) is used instead of {\Ne}({\cliii}), these {\te} values remain nearly identical. 
This consistency can be attributed to the fact that {\Ne}({\siii}) and {\Ne}({\cliii}) diagnose similar physical conditions, resulting in comparable outputs.

Both {\te}({\siii})${\rm opt}$ and {\te}({\siii})${\rm opt/ir}$ correspond to the intersection points of their respective {\Ne}--{\te} curves with the {\Ne}--{\te} curve derived from {\Ne}({\siii}). 
Although they represent the same ionization stage, the calculated {\te}({\siii}) values are slightly higher than those of {\te}({\oiii}), {\te}({\ariii}), and {\te}({\feiii}).
This difference is likely caused by uncertainties in the flux calibration of {\siii}\,9069\,{\AA} due to atmospheric extinction.

In the calculation of {\te}({\nii}), the recombination contribution of N$^{2+}$ to the {\nii}\,5755\,{\AA} line flux by following \citet{2000MNRAS.312..585L} is first subtracted.
The resulting {\te}({\nii}) value corresponds to the intersection point of its respective {\Ne}--{\te} curve with the one derived from {\Ne}({\sii}).

{\te}({\hei}) has been calculated using the flux ratio of selected pairs of {\hei} lines, such as {\hei} 
at 7281, 6678, and 4921\,{\AA} (singlet lines) or 5876\,{\AA} and 4471\,{\AA} (triplet lines). 
However, as this method is highly optimized for the specific line pairs chosen, the ionic He$^{+}$ abundances derived using {\te}({\hei}) and the {\hei} line fluxes occasionally exhibit scatter among the different lines. 
As demonstrated by \citet{2009ApJ...705..509O}, various {\hei} ratios produce slightly differing {\te}({\hei}) values. 
To address the He$^{+}$ abundance inconsistencies arising from {\te}({\hei}), we calculate a {\te}({\hei}) value that minimizes the scatter between the He$^{+}$ abundances obtained from individual {\hei} lines and the average He$^{+}$ abundance derived across all {\hei} lines.
In this process, we utilize the {\Ne}--{\te} logarithmic interpolation function of the {\hei} effective recombination coefficients by \citet{2022MNRAS.511.4774O} along with the {\Ne} diagnostics and the {\Ne} value determined from {\cliii}. 
Notably, even if {\Ne}({\sii}) is used instead of {\Ne}({\cliii}), the {\te}({\hei}) value remains essentially unchanged.

\subsection{Ionic abundances}

\begin{table}
\caption{Derived ionic abundances in DdDm\,1. \label{T-ionic}}
\begin{tabularx}{\columnwidth}{@{\extracolsep{\fill}}
l@{\hspace{5pt}}c@{\hspace{3pt}}D{p}{\pm}{-1}
@{}
}
\midrule	 
Ion X$^{i+}$  &Adopted {\te}/{\Ne} &\multicolumn{1}{c}{$n$(X$^{i+}$)/$n$(H$^{+}$)}\\
\midrule	 
He$^{+}$  (5)&{\te}({\hei})/{\Ne}({\cliii})  &9.11\times10^{-2}~p~6.61\times10^{-3} \\
C$^{2+}$  (1)&{\te}({\oiii})/{\Ne}({\cliii}) &9.48\times10^{-6}~p~1.40\times10^{-6} \\
N$^{0}$   (1)&{\te}({\nii})/{\Ne}({\sii})    &2.46\times10^{-7}~p~4.15\times10^{-8} \\
N$^{+}$   (3)&{\te}({\nii})/{\Ne}({\sii})    &4.80\times10^{-6}~p~1.88\times10^{-7} \\
N$^{2+}$  (1)&{\te}({\oiii})/{\Ne}({\cliii}) &1.48\times10^{-5}~p~2.74\times10^{-6} \\
O$^{0+}$  (2)&{\te}({\nii})/{\Ne}({\sii})    &1.53\times10^{-6}~p~8.32\times10^{-8} \\ 
O$^{1+}$  (2)&{\te}({\nii})/{\Ne}({\sii})    &2.68\times10^{-5}~p~5.84\times10^{-6} \\
O$^{2+}$  (4)&{\te}({\oiii})/{\Ne}({\cliii}) &8.45\times10^{-5}~p~1.18\times10^{-6} \\
Ne$^{+}$  (1)&{\te}({\nii})/{\Ne}({\sii})    &8.14\times10^{-6}~p~7.23\times10^{-7} \\
Ne$^{2+}$ (1)&{\te}({\oiii})/{\Ne}({\cliii}) &1.67\times10^{-5}~p~1.43\times10^{-6} \\
Si$^{2+}$ (2)&{\te}({\oiii})/{\Ne}({\cliii}) &1.47\times10^{-6}~p~2.07\times10^{-7} \\
S$^{+}$   (2)&{\te}({\nii})/{\Ne}({\sii})    &1.61\times10^{-7}~p~1.41\times10^{-8} \\
S$^{2+}$  (4)&{\te}({\siii})$_{\rm opt}$/{\Ne}({\siii}) &1.13\times10^{-6}~p~7.46\times10^{-8} \\ 
S$^{3+}$  (1)&{\te}({\siii})$_{\rm opt}$/{\Ne}({\siii}) &2.41\times10^{-7}~p~2.88\times10^{-8} \\  
Cl$^{2+}$ (2)&{\te}({\oiii})/{\Ne}({\cliii}) &1.64\times10^{-8}~p~8.42\times10^{-10} \\ 
Ar$^{1+}$ (1)&{\te}({\nii})/{\Ne}({\sii}) &6.56\times10^{-8}~p~6.07\times10^{-9} \\ 
Ar$^{2+}$ (3)&{\te}({\ariii})/{\Ne}({\cliii}) &4.38\times10^{-7}~p~8.31\times10^{-9} \\  
Ar$^{3+}$ (1)&{\te}({\ariii})/{\Ne}({\cliii}) &2.92\times10^{-8}~p~6.03\times10^{-9} \\  
Fe$^{2+}$ (5)&{\te}({\feiii})/{\Ne}({\cliii}) &6.37\times10^{-7}~p~1.07\times10^{-7} \\ 
\midrule
\end{tabularx}
\end{table}

The ionic abundances $n$(X$^{i+}$)/$n$(H$^{+}$) are summarized in table\,\ref{T-ionic}. 
The number in parentheses in the 1st column indicates the number of emission lines used for each ionic abundance calculation. 
The ionic abundances except for the He$^{+}$ abundance are determined by solving the population equations for multiple energy levels using the {\te} and {\Ne} values listed in the second column. 
When multiple lines are available, the ionic abundance is calculated for each line individually, followed by a three-sigma clipped averaging of the results. 
By following \citet{2000MNRAS.312..585L}, the respective O$^{+}$ and N$^{+}$ abundances derived from the {\oii}\,7320/7330\,{\AA} and {\nii}\,5755\,{\AA} lines are corrected for the recombination contribution from O$^{2+}$ and N$^{2+}$.
In calculating X$^{i+}$/H$^{+}$, the {\te} and {\Ne} values for H$^{+}$ are assumed to be the same as those for the co-existing X$^{i+}$.

\subsection{Elemental abundances \label{S:eabund1}}

\begin{table*}
\caption{Comparisons between elemental abundances in DdDm\,1 and the AGB nucleosynthesis result. \label{T-abund}}
\begin{tabularx}{\textwidth}{@{\extracolsep{\fill}}
l
@{\hspace{0pt}}D{p}{\pm}{-1}
@{\hspace{0pt}}D{p}{\pm}{-1}
@{\hspace{0pt}}D{.}{.}{-1}
@{\hspace{0pt}}D{p}{\pm}{-1}
@{\hspace{0pt}}D{p}{\pm}{-1}
@{\hspace{0pt}}D{.}{.}{-1}
@{\hspace{0pt}}D{p}{\pm}{-1}
@{}
}
\midrule
X  &\multicolumn{1}{c}{$n$(X)/$n$(H)} &\multicolumn{5}{c}{$12+\log_{10}\,$$n$(X)/$n$(H)}&\multicolumn{1}{c}{[X/H]} \\
\cline{3-7}
   &\multicolumn{1}{c}{this work} &\multicolumn{1}{c}{this work} &\multicolumn{1}{c}{Ref.(1)} &\multicolumn{1}{c}{Ref.(2)} &\multicolumn{1}{c}{Ref.(3)} 
&\multicolumn{1}{c}{AGB model}&\multicolumn{1}{c}{this work}\\
\midrule
He & 9.11\times10^{-2}~p~6.61\times10^{-3} & 10.96~p~0.03 & 10.95 & 10.97~p~0.06        & 11.01~p~0.04 & 10.98 & +0.03~p~0.03\\ 
C & 1.27\times10^{-5}~p~2.10\times10^{-6} & 7.10~p~0.07 & 7.05 & 7.05~p~0.20            & 7.02~p~0.06 & 7.66 & -1.33~p~0.09\\ 
N & 1.99\times10^{-5}~p~2.74\times10^{-6} & 7.30~p~0.06 & 7.40 & 7.40~p~0.11            & 7.36~p~0.02 & 7.47 & -0.53~p~0.08\\ 
O & 1.13\times10^{-4}~p~5.96\times10^{-6} & 8.05~p~0.02 & 8.06 & 8.06~p~0.09            & 8.07~p~0.01 & 8.01&-0.64~p~0.06 \\ 
Ne & 2.52\times10^{-5}~p~2.49\times10^{-6} & 7.40~p~0.04 & 7.24 & 7.32~p~0.08           & 7.45~p~0.01 &7.27 &-0.53~p~0.11 \\ 
Si & 1.96\times10^{-6}~p~3.13\times10^{-7} & 6.29~p~0.07 & \multicolumn{1}{r}{$\cdots$} & 6.25~p~0.13 & 6.58~p~0.15 &6.85& -1.22~p~0.08 \\
S & 1.53\times10^{-6}~p~2.53\times10^{-7} & 6.19~p~0.07 & 6.35 & 6.31~p~0.12 & 6.35~p~0.01 &6.46& -0.94~p~0.08 \\
Cl & 2.00\times10^{-8}~p~1.94\times10^{-9} & 4.30~p~0.04 & 4.69 & 4.42~p~0.10 & 4.61~p~0.08 &4.53& -1.20~p~0.30 \\
Ar & 5.33\times10^{-7}~p~3.10\times10^{-8} & 5.73~p~0.03 & 5.16 & 5.81~p~0.10 & 5.79~p~0.09 &5.74& -0.67~p~0.13 \\ 
Fe & 1.52\times10^{-6}~p~4.65\times10^{-7} & 6.18~p~0.13 & \multicolumn{1}{r}{$\cdots$} & \multicolumn{1}{r}{$\cdots$} & 6.58~p~0.14 &6.84 & -1.32~p~0.13\\ 
\midrule
\end{tabularx}
\begin{tabnote}
Note -- The C abundance are derived from CELs. Solar abundances are taken from \citet{Asplund:2009aa}. The AGB model result is taken from \citet{2018MNRAS.477..421K} 
where adopts the following values for the initial elemental abundances:  He:10.92, C:7.75, N:7.15, O:8.01, Ne:7.25, Si:6.83, S:6.44, Cl:4.51, Ar:5.72, Fe:6.82.\\
References -- 
(1) \citet{2005MNRAS.362..424W}; 
(2) \citet{2008ApJ...680.1162H};
(3) \citet{2009ApJ...705..509O}. 
\end{tabnote}
\end{table*}

Table\,\ref{T-abund} summarizes the derived elemental abundance $n$(X)/$n$(H) and $12+\log_{10}\,$$n$(X)/$n$(H). 
We emphasize that the values presented here are based on spectro-photometry of the entire nebula with absolutely no flux loss. 

The elemental abundances are calculated by introducing the ionization correction factor (ICF) for each element X. 
ICFs account for the ionic abundances of unobserved ionization stages within the observed spectral coverage. 
$n$(X)/$n$(H) can be expressed as ICF(X)$\cdot$$\sum n$(X$^{i+}$)/$n$(H$^{+}$)$_{\rm obs}$, where $n$(X$^{i+}$)/$n$(H$^{+}$)$_{\rm obs}$ 
refers to the ionic abundances derived from the observed spectra. 
The ICF values for He, N, O, S, and Ar are set to unity because the central star of DdDm\,1 has a low temperature, 
meaning the nebular He$^{2+}$, N$^{3+}$, O$^{3+}$, S$^{4+}$, and Ar$^{4+}$ abundances are expected to be negligible. 
For both C and Si, the ICF is given as $n$(O)/$n$(O$^{2+}$). For Ne, it is $n$(O)/($n$(O$^{+}$) + $n$(O$^{2+}$)). The ICF for Cl is $n$(Ar)/$n$(Ar$^{2+}$). 
For Fe, we adopt the ICF presented by \citet{2009ApJ...694.1335D}. 
The last column of table\,\ref{T-abund} shows the enhancement of each element with respect to solar abundances, where we refer to \citet{Asplund:2009aa}.

\section{Gas-to-Dust mass ratio, gas and dust masses}
\label{S-gdr}

We estimate the gas-to-dust mass ratio (GDR, $\psi$) and the gas/dust masses ($m_{\rm g}$/$m_{\rm d}$) following \citet{2022MNRAS.511.4774O}. 
When gas and dust grains coexist in the same volume, $\psi$ can be expressed as follows:

\begin{eqnarray}
\psi &=& \frac{C (\log_{10}\,e) \kappa_{{\rm H}\beta}}{c({\rm H}\beta)_{\rm cir}} \cdot \sum \frac{4 \pi I_{g} \mu_{g} }{n_{\rm e} j_{\rm g}(T_{\rm e},n_{\rm e}) },  \notag \\ 
\kappa_{{\rm H}\beta} &=& \frac{\int_{min}^{max} \pi a^{2}Q_{\rm ext}({\rm H}\beta,a) n(a) da}
{\int_{min}^{max} \frac{4}{3} \pi a^{3} \rho_{\rm d} n(a) da}, \label{eq-5} 
\end{eqnarray}

\noindent where $C$ is a dimensionless constant: assuming that 
the nebular shape is circular with a radius $r = 1${\arcsec} (subsection\,\ref{S:maps}) when projected on sky, and it equals $\pi^{-1}(4.848 \times 10^{-6}~r)^{-2}$. 
$\kappa_{{\rm H}\beta}$ is the mass extinction coefficient. 
$Q_{\rm ext}$({\hb},$a$) is the extinction coefficient of a grain with radius $a$ at {\hb}. 
$\rho_{\rm d}$ is the mass density of the grain (2.228\,g\,cm$^{-3}$). 
We adopt the optical constant of amorphous silicate grain taken from 
\cite{1993ApJ...402..441L}. 
$n(a)$ is the grain size distribution; we adopt $n(a) \propto a^{-3.5}$. 
We calculate the $\kappa_{{\rm H}\beta}$ value of $3.99\times10^{4}$\,cm$^{2}$\,g$^{-1}$ between the maximum radius 0.25\,{\micron} and the minimum radius 0.03\,{\micron} by following Mie theory.
$c$({\hb})$_{\rm cir}$ is the extinction coefficient at {\hb} stemming from circumstellar dust (i.e., $0.098 \pm 0.041$ estimated in section\,\ref{S-ext}).
$n{\rm g}$ and $\mu_{\rm g}$ denote the number density and atomic weight of the target gas, respectively.
$j_{\rm g}$({\te},{\Ne}) represents the volume emissivity of the gas component per $n_{\rm g}n_{\rm e}$.
$I_{\rm g}$ is the observed extinction-free line flux of the gas component.
In calculating GDR, we utilize the {\hi} and {\hei} lines.
For each ion, we adopt the same {\te} and {\Ne} values listed in table\,\ref{T-diagno}. 
Then, we obtain GDR of $619 \pm 102$.

Next, using equation\,(\ref{eq-6}), we directly calculate the total atomic gas mass $m_{\rm g}$
of $(1.26 \pm 0.15)\times10^{-3}$\,$D_{\rm kpc}^{2}$\,$f$\,M$_{\sun}$, where $D_{\rm kpc}$ 
is the distance in kpc and $f$ is a correction factor for the gas density assuming a uniform distribution (i.e., filling factor of the gas and dust emitting volume). 

\begin{eqnarray}
m_{\rm g} &=& \sum \frac{4 \pi D^{2} I_{g} \mu_{g}f}{n_{\rm e} j_{\rm g}(T_{\rm e},n_{\rm e}) }.
\label{eq-6}
\end{eqnarray}

Using the derived $m_{\rm g}$ and GDR values, we estimate the dust mass $m_{\rm d}$ to be 
$(4.05 \pm 0.86)\times10^{-6}$\,$D_{\rm kpc}^{2}$\,$f$\,M$_{\sun}$. 
We use equation (\ref{eq-6}) to avoid uncertainties related to the unknown geometric structure of the emitting region, under the assumption that the gas is uniformly distributed.

So far, the only report on GDR in DdDm\,1 comes from the pioneering work of \cite{1988MNRAS.235.1049H}. 
This study employed silicate dust radiative transfer models adopting $n(a) \propto a^{-3.5}$ with $a$ ranging from 0.005 to 0.25\,{\micron}, 
and obtained a GDR of 769 by fitting to the \emph{IRAS} four-band flux densities. 
However, the \emph{IRAS} 12\,{\micron} and 100\,{\micron} flux densities are upper limits, 
while the uncertainties for the 25\,{\micron} and 60\,{\micron} flux densities would be 30\,$\%$ or greater. 
Since silicate dust peaks at 9\,{\micron} and 18\,{\micron}, the uncertainties in the 12\,{\micron} and 25\,{\micron} data would undoubtedly have a significant impact on the model fitting. 
Thus, the GDR reported by \cite{1988MNRAS.235.1049H} likely includes an uncertainty of at least $\sim30$\,$\%$, leading to their GDR = $769 \pm 230$. 
If we adopt $a=0.005-0.25$\,{\micron} with $n(a) \propto a^{-3.5}$, we derive the GDR of $481 \pm 79$.  
In both $a$ ranges, our derived GDR values are consistent with that of \cite{1988MNRAS.235.1049H} within one-$\sigma$ uncertainty.

\section{Discussion}
\label{S-dis}

\subsection{Comparison of elemental abundances \label{S:eabund2}}

We compare our results for DdDm\,1 with three previously conducted observations in the 4-6th columns of table\,\ref{T-abund}. 
The used datasets are as follows: 
\citet{2008ApJ...680.1162H} utilized spectro-photometry obtained by a long-slit with a slit-width of 5{\arcsec} putting along P.A. = 90{\degree}, 
\citet{2009ApJ...705..509O} employed an echelle spectrum obtained with a slit-width of 0.6{\arcsec} and an ADC, 
and \citet{2005MNRAS.362..424W} used a long-slit spectrum obtained by a long-slit with a slit-width of 0.9{\arcsec} putting along P.A. = 0{\degree}, without an ADC. 
While our result is in excellent agreement with \citet{2008ApJ...680.1162H}, significant discrepancies of approximately 
a factor of two in the Cl abundance are seen when compared to the others. 
We examine the causes of these discrepancies.

First, we consider the discrepancy between our results and those of \citet{2009ApJ...705..509O}. 
This discrepancy likely arises from differences in the spectral extraction region and the adopted $c$({\hb}) value. The KOOLS-IFU spatially integrated 1-D spectra encompass the entire nebula, whereas \citet{2009ApJ...705..509O} focused on the bright central region.
The {\cliii} line maps generated from KOOLS-IFU reveal that their spatial distributions are concentrated in the central region and are more compact than that of the {\hb} line. 
Therefore, we speculate that the flux ratio of {\cliii} to {\hb} lines in \citet{2009ApJ...705..509O} is larger than our results.
To verify this hypothesis, we extracted a pseudo-spectrum from the KOOLS-IFU datacube, simulating the slit dimensions used by \citet{2009ApJ...705..509O}, and measured the flux ratio $F$({\cliii}\,5537\,{\AA})/$F$({\hb}). 
Our measurement reproduces their results well: $0.302 \pm 0.060$ in the pseudo-spectrum compared to $0.357 \pm 0.050$ reported by \citet{2009ApJ...705..509O}. 
Thus, we conclude that the differences in observational methods and spatial extraction regions are the primary reasons for the discrepancy in the chlorine abundance measurements.

Next, we address the discrepancy with \citet{2005MNRAS.362..424W}. 
While this discrepancy may partly arise from differences in the spectral extraction regions, a significant issue lies in the detection of the {\cliii}\,5537\,{\AA} line itself. 
They report detecting only the {\cliii}\,5537\,{\AA} line; however, this is highly unusual. 
This is because the {\cliii}\,5517\,{\AA} line flux is comparable to that of {\cliii}\,5537\,{\AA} line in DdDm\,1, as confirmed by all other observational results. 
In our observation, the ratio of {\cliii} $F$(5517\,{\AA}) to $F$(5537\,{\AA}) is $0.969 \pm 0.060$, which is consistent with previous measurements. 
The {\cliii} $F$(5517\,{\AA})/$F$(5537\,{\AA}) ratio indicates that both lines should be detectable with comparable fluxes.
Therefore, the reported detection of only the {\cliii}\,5537\,{\AA} line in \citet{2005MNRAS.362..424W} strongly suggests inaccuracies in their measurement, 
and the Cl abundance derived from their results cannot be considered reliable.

\citet{2005MNRAS.362..424W} did not calculate the Ne$^{+}$ abundance and adopted a higher {\te} value than ours for the Ar$^{2+}$ abundance derivation.
As a result, noticeable discrepancies in Ne and Ar abundances are found between \citet{2005MNRAS.362..424W} and ours.
The discrepancies in Si and Fe abundances between \citet{2009ApJ...705..509O} and our results are due to differences in the adopted ICFs: 
\citet{2009ApJ...705..509O} used the O/O$^{+}$ ratio, whereas we adopt the O/O$^{2+}$ ratio for Si and 1.1\,(O$^{+}$/O$^{2+}$)$^{0.58}$\,(O/O$^{+}$) from \citet{2009ApJ...694.1335D} for Fe.
Thus, we account for the abundance discrepancies observed between previous studies and our results.

In summary, our results align with previous observations and can be regarded as representative elemental abundance values for DdDm\,1.
Furthermore, we recognize that spatial coverage is critical for accurately estimating abundances in PNe.

\subsection{Comparison with AGB nucleosynthesis models}
\label{S:d-abund}

The most notable characteristics of DdDm\,1 are its extremely low carbon abundance and low metallicity. 
The [Ar/H] value (see the last column of table\,\ref{T-abund}) indicates that the metallicity of DdDm\,1 is comparable to that of the Small Magellanic Cloud (SMC).
Argon, a noble gas not synthesized in nucleosynthesis processes within progenitor stars, is resistant to depletion onto dust grains or molecules, making it a robust tracer of metallicity.
To the best of our knowledge, DdDm\,1 is the most carbon-deficient PN in the Milky Way. 
Carbon is synthesized during the AGB phase and transported to the stellar surface through the TDUs, a mixing process triggered by He-shell instabilities or thermal pulses. 

According to the theoretical AGB nucleosynthesis models by \citet{2018MNRAS.477..421K} (see table\,\ref{T-abund}), the efficiency of the TDUs increases significantly in stars with initial masses of $\sim1.15$\,M$_{\sun}$ or higher, particularly at the SMC metallicity. 
TDU events enrich the stellar surface with carbon and other elements, with this enrichment becoming more pronounced in stars with initial masses up to $\sim3-4$\,M$_{\sun}$. 
The extraordinarily low carbon abundance of DdDm\,1 strongly indicates that its progenitor star had an initial mass of less than 1.15\,M$_{\sun}$. 
In fact, the model by \citet{2018MNRAS.477..421K} predicts no TDU in such stars.
 Instead, it shows that during the thermal pulse AGB phase, the isotope ratios of C, N, and O undergo slight changes due to the effects of weak convection, the CNO cycle, or very weak and inefficient TDUs. 
These changes can explain the observed C and N abundances in DdDm\,1: C is slightly depleted, while N is slightly enhanced relative to their initial values. 
Similarly, the absence of significant surface enrichment of Ne as well as C and O further supports the conclusion that DdDm\,1 did not experience efficient TDU events. 
Nitrogen is typically enhanced through the CNO cycle during earlier phases, such as the red giant branch (RGB) phase, and/or through the hot bottom burning (HBB) process during the late AGB phase. 
However, the HBB process only occurs in stars with initial masses of $\sim3$\,M$_{\sun}$ or higher, ruling out its contribution to DdDm\,1. 
While uncertainties in the derived carbon abundance—such as potential observational errors or model assumptions—should be considered, the observed elemental abundances in DdDm\,1 are consistent with the predictions of the AGB model 
for a star with an initial mass of 1.0\,M$_{\sun}$ by \citet{2018MNRAS.477..421K}. 
To the best of our knowledge, this is the first evidence showing that both the model and observations conclusively rule out TDUs in DdDm\,1.

The lower [Si,Fe/H] values compared to the [Ar/H] values suggest that both elements are likely sequestered in dust grains, consistent with their depletion in the gas phase.

In conclusion, DdDm\,1 most likely evolved from a progenitor star with an initial mass of $\sim1.0$\,M$_{\sun}$. 
The observed carbon abundance suggests that the progenitor star formed in an extremely carbon-poor environment and evolved as a single star, without undergoing mass accretion events such as binary mass transfer or coalescence during its evolution. 
DdDm\,1 did not experience efficient TDUs.
Furthermore, the nebular abundances appear to closely reflect the initial composition of the progenitor star, making DdDm\,1 a particularly interesting intriguing PN. 
It provides observational evidence that PNe can form and evolve even from stars with such low initial masses.

\subsection{Post-AGB evolution of the central star}
\label{S:d-evolution}

\begin{figure}
\includegraphics[width=\columnwidth]{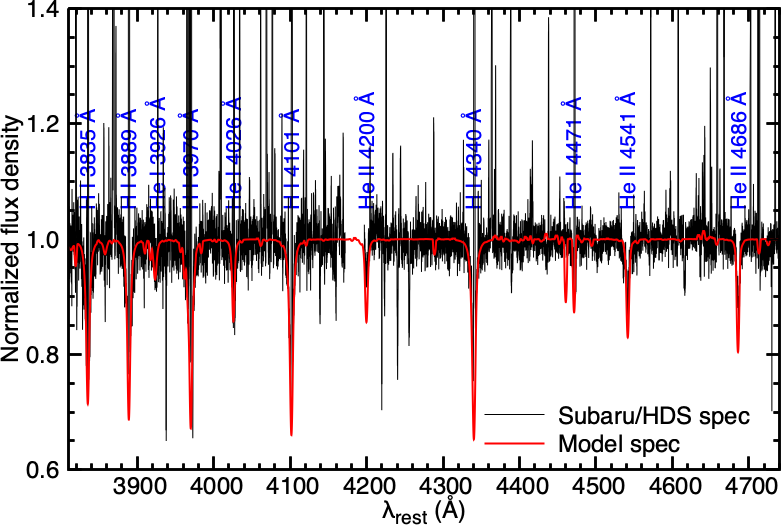}
\caption{
Comparison between the Subaru/HDS spectrum of the central star and its non-LTE theoretical spectrum synthesized with the {\sc Tlusty} code \citep{1988CoPhC..52..103H}.  
In the fitting procedure, the nebular elemental abundances (see table\,\ref{T-abund}), $T_{\rm eff}$ of 41639\,K, and $\log_{10}\,g$ of 3.93\,cm\,s$^{-2}$ derived from our photoionization model 
(see table\,\ref{T-model} in subsection\,\ref{S:d-phot}) are adopted.
\label{F-subaru}
{Alt text: Comparison between the Subaru/HDS spectrum of the central star 
and its theoretically calculated spectrum. The horizontal and vertical axes are wavelength and normalized flux density, respectively.}
}
\end{figure}

We infer the post-AGB evolution of the central star. 
\citet{2008ApJ...680.1162H} suggest that the central star is a He-shell burner. 
However, the Subaru HDS spectrum of the central star (figure\,\ref{F-subaru}) clearly exhibits the stellar absorption lines attributed to {\hi}, {\hei}, and {\heii}. 
Thus, we assume that the H-shell burning can sustain the luminosity of the central star after the AGB phase. 

We synthesize a post-AGB evolution track for a star with an initial mass of 1.0\,M$_{\sun}$ and $Z = 0.0028$, 
corresponding to SMC metallicity. As presented in figure\,\ref{F-sed} (c) and (d), we create the evolution track by interpolating between 
the theoretical post-AGB evolution tracks for 1.0\,M$_{\sun}$ stars with $Z$ = 0.010 and 0.001 provided by \citet{2016A&A...588A..25M}. 
Previous studies have reported $T_{\rm eff}$ values ranging from 37000 to 55000\,K \citep[see table\,1 of][]{2009ApJ...705..509O};
Based on this, two possible evolutionary statuses can be considered: (1) the central star is still evolving toward the whitedwarf (WD) cooling track, 
with its temperature increasing; or (2) the central star has already entered the WD cooling track, with its temperature decreasing.

The nebular spectrum of DdDm\,1 shows neither {\heii} nebular emission lines nor mid-infrared molecular hydrogen (H$_{2}$) emission.
Additionally, we confirm that the central star spectrum obtained with Subaru HDS displays narrow absorption lines, indicating low surface gravity.
Consistently, \citet{1997IAUS..180..120M} directly measured $\log_{10}\,g$ = 3.4\,cm\,s$^{-2}$ using a Keck HIRES spectrum of DdDm\,1. 
These characteristics are inconsistent with those typically observed in PNe whose central stars have already entered the WD cooling track. 
Most of PNe in the WD cooling track show {\heii} nebular emission lines, their central star spectra exhibit broad absorption lines.

If the central star is in the WD cooling track with $T_{\rm eff} \le 46000$\,K, 
$L_{\ast}$ and $\log_{10}\,g$ would be $\le 1.25$\,L$_{\sun}$ and $\ge 7.7$\,cm\,s$^{-2}$, respectively, 
according to our synthesized evolution track for a star with initially 1.0\,M$_{\sun}$ and $Z = 0.0028$. 
Using the distance estimate derived from our luminosity budget method (subsection\,\ref{S:d-filling}), the distance to DdDm\,1 would then be $\le 0.42$\,kpc.
At this distance, the {\it outer} radius of the nebula would correspond to only $\le 0.002$\,pc, which is unacceptably small and physically implausible. 
Furthermore, $\log_{10}\,g \ge 7.7$\,cm\,s$^{-2}$ cannot explain the absorption profile of the observed spectrum of the central star at all.

The central star of DdDm\,1 did not evolve as a hydrogen burner but may have undergone a born-again event during the post-AGB phase, similar to Hen\,3-1357 \citep[the Stingray Nebula; see][for details]{2017ApJ...838...71O}. 
Hen\,3-1357 is an O-rich PN that exhibits 9/18\,{\micron} silicate emission in its \emph{Spitzer}/IRS spectrum and is shows short-term variations in $\log_{10}\,g$, $T_{\rm eff}$, and $L_{\ast}$ 
\citep{1993A&A...267L..19P,2014A&A...565A..40R,2017MNRAS.464L..51R,2021ApJ...907..104B,2022MNRAS.515.1459P}.
However, no evidence of such variations, such as changes in surface gravity, effective temperature, or luminosity, has been observed in DdDm 1, making this scenario unlikely.

In summary, the central star of DdDm\,1 as a H-burner is currently evolving toward the WD cooling track, with its temperature increasing while maintaining nearly constant luminosity.

\subsection{Ejected total gas mass}
\label{S:d-gasmas} 

In the synthesized evolutionary tracks, $L_{\ast}$ and $\log_{10}\,g$ are established as a function of $T_{\rm eff}$ for each. 
For a $T_{\rm eff}$ range of $37000-55000$\,K, the corresponding ranges for $L_{\ast}$ and $\log_{10}\,g$ are $4725-4877$\,L$_{\sun}$ and $3.72-4.43$\,cm\,s$^{-2}$, respectively. 
\citet{2018MNRAS.477..421K} predict that a star with an initial mass of 1.0\,M$_{\sun}$ and $Z=0.0028$ will evolve into a central star with a final mass of 0.651\,M$_{\sun}$ after ejecting 0.349\,M$_{\sun}$. 
Similarly, \citet{2016A&A...588A..25M} predict that stars with an initial mass of 1.0\,M$_{\sun}$ and $Z=0.001$ and $Z=0.010$ will evolve into central stars with core masses of $\sim0.55$\,M$_{\sun}$ and $\sim0.53$\,M$_{\sun}$, respectively, implying mass ejection of $\sim0.45-0.47$\,M$_{\sun}$. 

In summary, stars including the progenitor of DdDm\,1 with an initial mass of 1.0\,M$_{\sun}$ and $Z=0.0028$ are expected to eject $\sim0.35-0.47$\,M$_{\sun}$ during their evolution. 
These predicted values - $T_{\rm eff}$, $L_{\ast}$, $\log_{10}\,g$, and the ejected mass - will serve as constraints for the 
photoionization modeling (subsection\,\ref{S:d-phot}).

\subsection{Emitting volume of gas and dust emission}
\label{S:d-filling} 

The nebular luminosity from ionized and atomic gas and dust grains should be in principle equal to the luminosity of the central star. 
If the nebular luminosity is less than that of the central star, it suggests that the ionization boundary is determined by the material distribution, the filling factor is less than unity, or possibly both. \citet{2009ApJ...705..509O} assumed that the nebula is material-bounded to reproduce the observed line fluxes in their photoionization model adopting $D = 10$\,kpc and $T_{\rm eff} = 39000$\,K, 
$L_{\ast} = 2000 - 3000$\,L$_{\sun}$, $\log_{10}\,g = 5.0$\,cm\,s$^{-2}$, and $f=1.0$. 
In the case of DdDm\,1, we think that the nebular boundary is determined not by the material distribution but by the radiation field of the central star because the KOOLS-IFU spectrum exhibits neutral atomic emission lines such as {\oi}.

Here, we assume that filling factor can be expressed as the ratio of the nebular luminosity to the central star luminosity. 
This means $f$ can be expressed as a function of $D_{\rm kpc}$ and $L_{\ast}$. 
We obtain the nebular luminosity using 
(1) a theoretically synthesized nebular continuum (2.22\,$D_{\rm kpc}^{2}$\,L$_{\sun}$, integrated in $0.1-10^{4}$\,{\micron}), 
(2) a modified two-temperature blackbody fitting (60.3\,K and 117.6\,K) with the grain-size weight-average $Q_{\rm ext}$($\lambda$,$a$) 
(for $a=0.03-0.25$\,{\micron} and an $a^{-3.5}$ size distribution) applied to the \emph{Spitzer} IRS spectrum (1.61\,$D_{\rm kpc}^{2}$\,L$_{\sun}$), 
and (3) theoretical calculations of emission line fluxes (3.59\,$D_{\rm kpc}^{2}$\,L$_{\sun}$), where 
we consider {\hi} ($2-1$, $n-2$, and $n-3$ transitions, where $n$ is the principal quantum number, up to $n=40$), 
{\hei}, $[${\ciii}$]$ + {\ciii}$]$, [N\,{\sc i, ii, iii}], [O\,{\sc i, ii, iii}], [Ne\,{\sc ii, iii}], [Si\,{\sc iii}], [S\,{\sc ii, iii, iv}], [Cl\,{\sc iii}], and [Ar\,{\sc ii, iii, iv}]. 

Thus, in the range of $T_{\rm eff} = 37000-55000$\,K, we find $f$ is $\sim0.35-0.60$ in $L_{\ast}$ = $4725-4877$\,L$_{\sun}$ and $D = 15-22$\,kpc. 
Especially, adopting $D=20$\,kpc (subsection\,\ref{S:maps}) and $L_{\ast} = 4800$\,L$_{\sun}$ leads to $f = 0.62$, and we obtain $m_{\rm g} = (0.31 \pm 0.04)$\,M$_{\sun}$ and 
using equation (\ref{eq-6}) and $m_{\rm d} = (1.00 \pm 0.21)\times10^{-3}$\,M$_{\sun}$ using equation (\ref{eq-5}). 
This $m_{\rm g}$ value satisfies with the theoretically predicted total ejected mass $\sim0.35-0.47$\,M$_{\sun}$ (subsection\,\ref{S:d-gasmas}).

\subsection{Photoionization modeling}
\label{S:d-phot}

\begin{table}
\caption{
\label{T-model}
The adopted and derived parameters in the {\sc Cloudy} model.
}
\begin{tabularx}{\columnwidth}{@{\extracolsep{\fill}}ll}
\midrule
Central star &Value \\ 
\midrule
$T_{\rm eff}$ &41639\,K\\
$L_{\ast}$ & 4848\,L$_{\sun}$\\ 
$\log_{10}\,g$&3.93\,cm\,s$^{-2}$\\
$D$ &19.39\,kpc\\
\midrule
Nebula &Value \\
\midrule
elemental abundances            &He: 10.98,  C: 6.92,  N: 7.37,  O: 8.06, \\
(12 + $\log_{10}$$n$(X)/$n$(H)) &Ne:  7.53, Si: 6.23,  S: 6.26, Cl: 4.24, \\ 
                                &Ar:  5.71, Fe: 6.37, \\
                                &others: \citet{2018MNRAS.477..421K}\\
geometry                        &spherical shell nebula\\
                                &inner radius = 47.3\,au\\ 
                                &ionization boundary radius = 19742\,au\\
                                &(=1.0\,arcsec)\\ 
$n(r)$  &4486\,cm$^{-3}$\\
filling factor                  &0.510\\
$I$({\hb})                      &$2.55\times10^{-12}$\,erg\,cm$^{-2}$\,s$^{-1}$\\
total gas mass                  &0.287\,M$_{\sun}$\\
\midrule
Silicate dust grain &Value \\
\midrule
temperature                     &$50-1230$\,K\\
total dust mass                 &$3.78\times10^{-4}$\,M$_{\sun}$\\
GDR                             &760\\
\midrule
Reduced-$\chi^{2}$ &22.01\\
\midrule
\end{tabularx}
\end{table}

We construct a photoionization model to reproduce all observed quantities (line fluxes, flux densities, and nebular radius) 
and the derived physical parameters, ensuring consistency with the post-AGB evolutionary model for stars with initially 1\,M$_{\sun}$ and $Z=0.0028$, within the measurement uncertainties. 
We use the photoionization code {\sc Cloudy} \citep[v17.03;][]{2017RMxAA..53..385F}.

\subsubsection{Modeling approach}

As discussed in subsection\,\ref{S:d-evolution}, the central star is currently evolving toward a white dwarf with increasing effective temperature. 
Referring to previous works, $T_{\rm eff}$ is estimated to range from 37000\,K to 55000\,K, and $\log_{10}\,g$ is $\sim 4.0$\,cm\,s$^{-2}$. 
Hence, as the incident SED, we adopt theoretical spectra of non-LTE line-blanketed model atmospheres of O-type stars by \citet{2003ApJS..146..417L}. 
This grid model is suitable for DdDm\,1 in the ranges of $T_{\rm eff}$ and $\log_{10}\,g$ ranges.
We set the photospheric elemental abundances to [X/H] = $-0.70$, except for He (He/H=0.10). 
We have established two distinct functions: one for $L_{\ast}$ and another for $\log_{10}\,g$, both are as functions of $T_{\rm eff}$.
Given a $T_{\rm eff}$ value, $L_{\ast}$ and $\log_{10}\,g$ are uniquely determined along the predicted post-AGB evolutionary track for 1\,M$_{\sun}$ stars with $Z=0.0028$.
We treat the distance $D$ as a free parameter, allowing it to vary within the range of $15-22$\,kpc.

We adopt a spherically symmetric nebula with a constant hydrogen radial profile $n(r)$ as a function of $r$, the distance from the central star and 
the ionization boundary is determined by the radiation from the central star. 
Given $L_{\ast}$ and $D$ values, the filling factor $f$ is uniquely determined using the observed nebular luminosity, as explained in subsection\,\ref{S:d-filling}.

We begin the modeling with the empirically determined elemental abundances (see table\,\ref{T-abund}) as input values. 
These abundances are then varied within their three-$\sigma$ uncertainties to achieve the best-fit model. 
Throughout the iterative process, the abundances of undetected elements lighter than zinc are fixed to the values predicted by 
\citet{2018MNRAS.477..421K} for 1.0\,M$_{\sun}$ stars with $Z=0.0028$.

We adopt spherical amorphous silicate grain with $a = 0.03-0.25$\,{\micron} following $a^{-3.5}$ size distribution.
We adopt the optical constant of \cite{1993ApJ...402..441L}. 
The radius range is divided into 30 bins. 
For each bin, {\sc Cloudy} calculates the absorption efficiency $Q_{\rm abs}$ and scattering efficiency $Q_{\rm sca}$ based on Mie theory. 
The resulting $Q_{\rm abs}$ and $Q_{\rm sca}$ values are then used in the model calculations.

\begin{figure*}
\includegraphics[width=\textwidth]{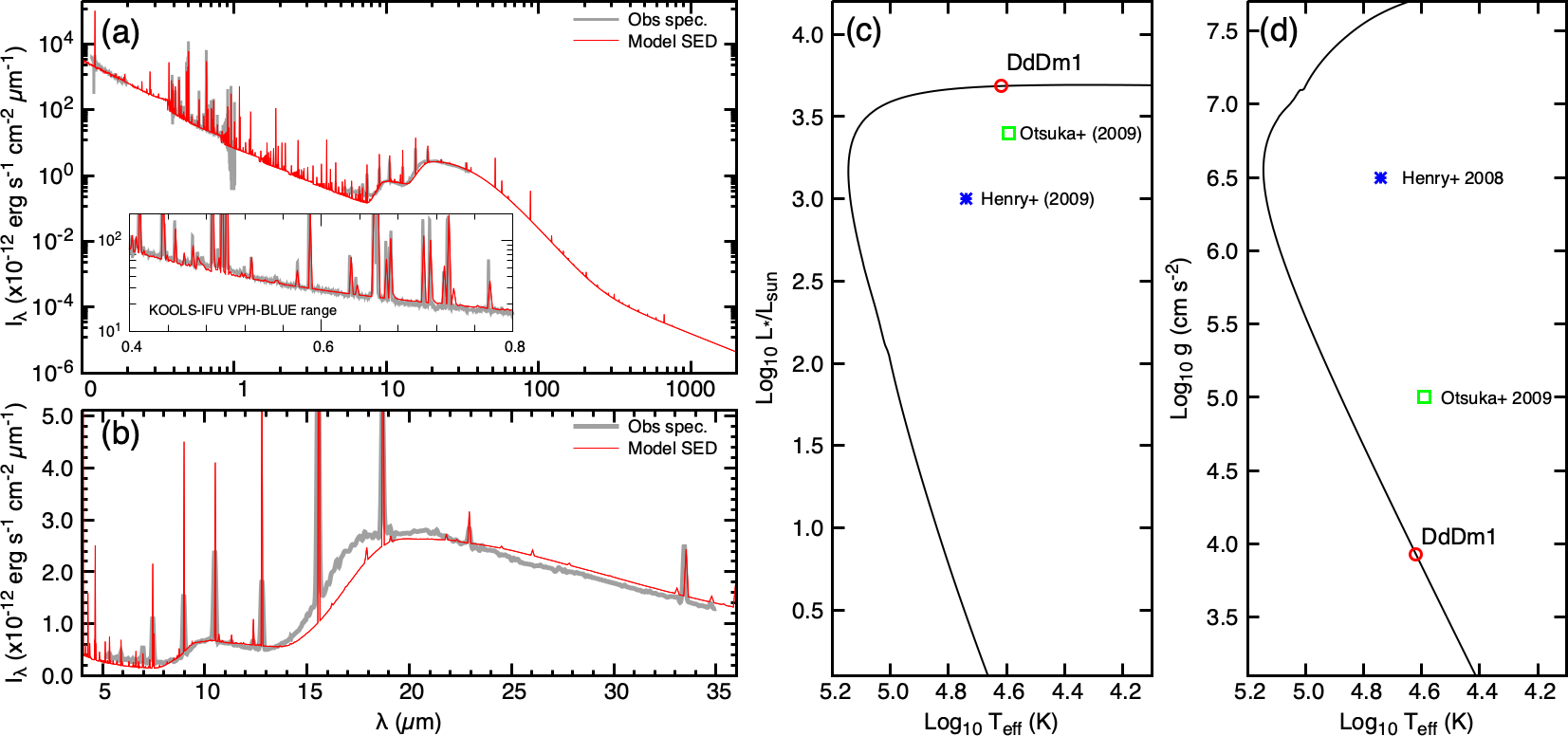}
\caption{
({\it left two panels}) Comparison between the observed spectra and the predicted SED by the best model. 
Panel (a) displays the $0.1-2000$\,{\micron} range SED. The inset focuses on the $0.4-0.8$\,{\micron} range. 
Panel (b) focuses on the $5-40$\,{\micron} range.
({\it right two panels}) The current location of the central star predicted by {\sc Cloudy} modeling is 
plotted on the $\log_{10}\,T_{\rm eff} - \log_{10}\,L_{\ast}$/L$_{\sun}$ plane (panel c) and the $\log_{10}\,T_{\rm eff} - \log_{10}\,g$ plane (panel d). 
The theoretical post-AGB evolutionary track for 1.0\,M$_{\sun}$ stars with $Z=0.0028$ (black line, synthesized using the data from \citet{2016A&A...588A..25M}) is plotted on both planes.
The values reported by \citet[Henry+ 2008]{2008ApJ...680.1162H} and \citet[Otsuka+ 2009]{2009ApJ...705..509O} are also included.
{Alt text: The left two panels display a comparison of the observed spectra with the theoretically calculated spectrum. 
The right two panels display the location of the central star on the theoretically predicted $\log_{10}\,T_{\rm eff} - \log_{10}\,L_{\ast}$/L$_{\sun}$ and $\log{10}\,T_{\rm eff} - \log_{10}\,g$ planes. 
Both the theoretical post-AGB evolutionary track and the values from the two prior studies are plotted on these planes.}
\label{F-sed}
}
\end{figure*}

We vary the following 14 free parameters until the model achieves the best: $T_{\rm eff}$, $D$, the inner radius of the nebula, GDR, and 10 elemental abundances (He/C/N/O/Ne/Si/S/Cl/Ar/Fe).
Iterative calculations are terminated either when the nebular gas temperature drops below 4000\,K or when the predicted total gas mass to 0.47\,M$_{\sun}$ (subsection\,\ref{S:d-gasmas}).
The quality of the fit is assessed based on the reduced-$\chi^{2}$ value, calculated from 97 observational constraints:
the {\hb} flux $I$({\hb}) (observed value: ($2.54 \pm 0.08$)$\times10^{-12}$\,{\ergsF}, see section\,\ref{S-ext}), 69 gas emission line fluxes relative to $I$({\hb}), 
25 mid-infrared \emph{Spitzer}/IRS flux densities, GDR (observed value: $619 \pm 102$, see subsection\,\ref{S-gdr}), 
and the ionization boundary radius of 1.0{\arcsec} (subsection\,\ref{S:maps}). 
In the reduced-$\chi^{2}$ calculation, we exclude the {\hei}\,7065\,{\AA} line because its reduced-$\chi^{2}$ value is large (708): 
the enhancement of collisional excitation for this line is important. 
Except for the {\oi} and {\NI} lines, we adopt the one-$\sigma$ uncertainty. For the {\oi} and {NI} lines, we set their uncertainty to 30\,$\%$.
Referring to the \emph{Spitzer}/IRS handbook, we adopt a one-$\sigma$ uncertainty of 6\,$\%$ for Band IRS01-09, which are covered by the SL module, and 10\,$\%$ for the other bands, 
which are covered by the SH and LH modules (see table\,\ref{T-line}). 
For the ionization boundary radius (i.e., the outer radius of the nebula), we adopt the one-$\sigma$ uncertainty of 30\,$\%$.

\subsubsection{Model results}

\begin{table*}
\caption{Comparison between the observed and the CLOUDY model predicted values.
\label{T-line}
}
\begin{tabularx}{\textwidth}{@{\extracolsep{\fill}}
l
c
@{\hspace{0pt}}D{.}{.}{-1}
@{\hspace{0pt}}D{.}{.}{-1}
@{\hspace{0pt}}D{.}{.}{-1}
l
c
@{\hspace{0pt}}D{.}{.}{-1}
@{\hspace{0pt}}D{.}{.}{-1}
@{\hspace{0pt}}D{.}{.}{-1}
@{}
}
\midrule
\multicolumn{1}{l}{\hspace{-6pt}Species} &
\multicolumn{1}{c}{$\lambda$} &
\multicolumn{1}{c}{$I$(obs)} &
\multicolumn{1}{c}{${\delta}I$(obs)} &
\multicolumn{1}{c}{$I$({\sc Cloudy})} &
\multicolumn{1}{l}{Species} &
\multicolumn{1}{c}{$\lambda$} &
\multicolumn{1}{c}{$I$(obs)} &
\multicolumn{1}{c}{${\delta}I$(obs)} &
\multicolumn{1}{c}{$I$({\sc Cloudy})} \\
\cline{3-5}
\cline{8-10}
 &
 &
\multicolumn{3}{c}{$I$({\hb})=100.000} &
&
&
\multicolumn{3}{c}{$I$({\hb})=100.000} \\
\midrule
O\,{\sc iii}$]$ & 1666.15\,{\AA} & 7.070 & 1.082 & 6.625 & {\sii} & 6716.44\,{\AA} & 2.733 & 0.101 & 2.733 \\ 
{\niii} & 1744-54\,{\AA} & 4.432 & 0.798 & 6.604 & {\sii} & 6730.82\,{\AA} & 4.630 & 0.106 & 4.829 \\ 
{\sIiii} & 1882.71\,{\AA} & 4.455 & 0.993 & 6.354 & {\hei} & 7065.22\,{\AA} & 6.403 & 0.109 & 9.300 \\ 
{\sIiii} & 1892.03\,{\AA} & 5.504 & 0.892 & 5.122 & {\ariii} & 7135.79\,{\AA} & 6.463 & 0.123 & 8.205 \\ 
$[${\ciii}$]$+{\ciii}$]$ & 1906/09\,{\AA} & 13.374 & 1.912 & 13.434 & {\hei} & 7281.35\,{\AA} & 1.105 & 0.059 & 1.015 \\ 
{\hi} & 4101.73\,{\AA} & 24.296 & 1.385 & 26.344 & {\oii} & 7318/19\,{\AA} & 6.234 & 0.249 & 8.865 \\ 
{\hi} & 4340.46\,{\AA} & 44.855 & 0.807 & 47.498 & {\oii} & 7330/31\,{\AA} & 6.576 & 0.250 & 7.113 \\ 
{\oiii} & 4363.00\,{\AA} & 4.758 & 0.366 & 5.546 & {\ariii} & 7751.11\,{\AA} & 1.690 & 0.066 & 1.968 \\ 
{\hei} & 4471.49\,{\AA} & 5.188 & 0.275 & 5.026 & {\hi} & 8358.96\,{\AA} & 0.258 & 0.055 & 0.178 \\ 
{\feiii} & 4658.01\,{\AA} & 2.390 & 0.105 & 2.413 & {\hi} & 8374.43\,{\AA} & 0.236 & 0.047 & 0.202 \\ 
{\feiii} & 4701.62\,{\AA} & 0.775 & 0.062 & 0.902 & {\hi} & 8392.35\,{\AA} & 0.282 & 0.032 & 0.231 \\ 
{\ariv}+{\hei} &4711/13\,{\AA}  & 0.882 & 0.039 & 1.037 & {\hi} & 8413.28\,{\AA} & 0.290 & 0.064 & 0.266 \\ 
{\feiii} & 4733.84\,{\AA} & 0.331 & 0.029 & 0.314 & {\hi} & 8437.91\,{\AA} & 0.467 & 0.064 & 0.310 \\ 
{\ariv} & 4740.12\,{\AA} & 0.146 & 0.029 & 0.189 & {\hi} & 8467.21\,{\AA} & 0.421 & 0.022 & 0.365 \\ 
{\feiii} & 4754.64\,{\AA} & 0.513 & 0.023 & 0.441 & {\hi} & 8502.44\,{\AA} & 0.482 & 0.026 & 0.435 \\ 
{\feiii} & 4769.52\,{\AA} & 0.328 & 0.024 & 0.302 & {\hi} & 8545.34\,{\AA} & 0.547 & 0.026 & 0.526 \\ 
{\feiii} & 4777.61\,{\AA} & 0.162 & 0.023 & 0.149 & {\hi} & 8598.35\,{\AA} & 0.698 & 0.044 & 0.645 \\ 
{\hi} & 4861.33\,{\AA} & 100.000 & 1.000 & 100.000 & {\hi} & 8664.98\,{\AA} & 0.803 & 0.053 & 0.805 \\ 
{\feiii} & 4881.12\,{\AA} & 1.225 & 0.061 & 1.035 & {\hi} & 8750.43\,{\AA} & 0.985 & 0.041 & 1.023 \\ 
{\hei} & 4921.93\,{\AA} & 1.204 & 0.060 & 1.306 & {\hi} & 8862.74\,{\AA} & 1.304 & 0.064 & 1.330 \\ 
{\oiii} & 4958.91\,{\AA} & 146.137 & 2.046 & 145.454 & {\hi} & 9014.87\,{\AA} & 1.999 & 0.116 & 1.775 \\ 
{\oiii} & 5006.84\,{\AA} & 433.515 & 6.069 & 433.973 & {\siii} & 9068.62\,{\AA} & 12.643 & 0.455 & 15.019 \\ 
{\feii} & 5158.78\,{\AA} & 0.182 & 0.037 & 0.261 & {\hi} & 9228.97\,{\AA} & 2.557 & 0.138 & 2.448 \\ 
{\NI} & 5199/5201\,{\AA} & 0.492 & 0.148 & 0.174 & {\hi} & 10049.3\,{\AA} & 5.626 & 0.225 & 5.315 \\ 
{\feiii} & 5270.40\,{\AA} & 1.197 & 0.087 & 1.848 & {\arii} & 6.98\,{\micron} & 0.928 & 0.082 & 0.259 \\ 
{\cliii} & 5517.71\,{\AA} & 0.200 & 0.011 & 0.163 & {\hi} & 7.46\,{\micron} & 3.316 & 0.295 & 1.959 \\ 
{\cliii} & 5537.87\,{\AA} & 0.206 & 0.008 & 0.226 & {\ariii} & 8.99\,{\micron} & 4.730 & 0.402 & 5.113 \\ 
{\nii} & 5755.00\,{\AA} & 1.269 & 0.037 & 1.248 & {\siv} & 10.51\,{\micron} & 9.170 & 0.779 & 5.260 \\ 
{\hei} & 5875.64\,{\AA} & 13.486 & 0.216 & 14.852 & {\hi} & 11.31\,{\micron} & 0.225 & 0.036 & 0.237 \\ 
{\oi} & 6300.30\,{\AA} & 1.982 & 0.595 & 1.881 & {\hi} & 12.37\,{\micron} & 0.716 & 0.068 & 0.719 \\ 
{\siii} & 6312.06\,{\AA} & 1.476 & 0.063 & 1.600 & {\neii} & 12.81\,{\micron} & 7.173 & 0.610 & 9.642 \\ 
{\oi} & 6363.78\,{\AA} & 0.695 & 0.209 & 0.601 & {\neiii} & 15.55\,{\micron} & 27.673 & 2.325 & 29.057 \\ 
{\nii} & 6548.05\,{\AA} & 15.071 & 2.788 & 15.853 & {\siii} & 18.71\,{\micron} & 13.778 & 1.171 & 14.627 \\ 
{\hi}                 & 6562.81\,{\AA} & 266.334 & 8.789 & 277.210 & {\feiii} & 22.92\,{\micron} & 1.855 & 0.326 & 1.905 \\ 
{\nii} & 6583.45\,{\AA} & 46.602 & 1.351 & 46.732 & {\siii} & 33.47\,{\micron} & 6.196 & 0.527 & 4.285 \\ 
{\hei} & 6678.15\,{\AA} & 3.517 & 0.070 & 3.681 &  &  &  &  &  \\ 
\midrule
\multicolumn{1}{l}{\hspace{-6pt}Band} &
\multicolumn{1}{c}{$\lambda$} &
\multicolumn{1}{c}{$F_{\nu}$(obs)} &
\multicolumn{1}{c}{${\delta}F_{\nu}$(obs)} &
\multicolumn{1}{c}{$F_{\nu}$({\sc Cloudy})} &
\multicolumn{1}{l}{Band} &
\multicolumn{1}{c}{$\lambda$} &
\multicolumn{1}{c}{$F_{\nu}$(obs)} &
\multicolumn{1}{c}{${\delta}F_{\nu}$(obs)} &
\multicolumn{1}{c}{$F_{\nu}$({\sc Cloudy})} \\
\cline{3-5}
\cline{8-10}
 &
 &
\multicolumn{3}{c}{(mJy)} &
&
&
\multicolumn{3}{c}{(mJy)} \\
\midrule
IRS01 & 5.00\,{\micron} & 2.992 & 0.180 & 2.201 & IRS14 & 19.00\,{\micron} & 334.000 & 33.400 & 309.500 \\ 
IRS02 & 6.00\,{\micron} & 3.871 & 0.232 & 2.321 & IRS15 & 20.00\,{\micron} & 364.600 & 36.460 & 351.700 \\ 
IRS03 & 7.00\,{\micron} & 4.673 & 0.280 & 2.560 & IRS16 & 21.00\,{\micron} & 396.100 & 39.610 & 387.200 \\ 
IRS04 & 8.00\,{\micron} & 5.148 & 0.309 & 4.108 & IRS17 & 22.00\,{\micron} & 426.400 & 42.640 & 421.800 \\ 
IRS05 & 9.00\,{\micron} & 11.940 & 0.716 & 12.380 & IRS18 & 23.00\,{\micron} & 446.600 & 44.660 & 449.700 \\ 
IRS06 & 10.00\,{\micron} & 20.080 & 2.008 & 22.070 & IRS19 & 24.00\,{\micron} & 458.500 & 45.850 & 478.200 \\ 
IRS07 & 11.00\,{\micron} & 25.610 & 2.561 & 25.870 & IRS20 & 25.00\,{\micron} & 468.300 & 46.830 & 502.600 \\ 
IRS08 & 12.00\,{\micron} & 26.110 & 2.611 & 29.200 & IRS21 & 26.00\,{\micron} & 480.800 & 48.080 & 522.200 \\ 
IRS09 & 13.00\,{\micron} & 26.000 & 2.600 & 32.200 & IRS22 & 27.00\,{\micron} & 499.300 & 49.930 & 538.500 \\ 
IRS10 & 15.00\,{\micron} & 84.760 & 8.476 & 61.960 & IRS23 & 28.00\,{\micron} & 519.400 & 51.940 & 551.400 \\ 
IRS11 & 16.00\,{\micron} & 150.100 & 15.010 & 105.100 & IRS24 & 29.00\,{\micron} & 530.900 & 53.090 & 562.400 \\ 
IRS12 & 17.00\,{\micron} & 224.900 & 22.490 & 168.400 & IRS25 & 30.00\,{\micron} & 533.500 & 53.350 & 568.900 \\ 
IRS13 & 18.00\,{\micron} & 291.100 & 29.110 & 241.900 &  &  &  &  &  \\ 
\midrule
\end{tabularx}
\end{table*}

Table\,\ref{T-model} summarizes all the input parameters and the derived quantities in the best model. 
The dust grain temperature peaks at $a=0.03$\,{\micron} in the innermost layer of the nebula (1230\,K) 
and reaches its minimum at $a=0.25$\,{\micron} in the outermost layer (50\,K).
Table\,\ref{T-line} provides a detailed comparison between the observed and model-derived line fluxes and dust continuum flux density. 
The reduced-$\chi^{2}$ value of 22.01 is comparable to that achieved in the model of H4-1 based on KOOLS-IFU data \citep{2023PASJ...75.1280O}.
Thus, we conclude that our model reproduces these observed quantities within acceptable accuracy.
Figure\,\ref{F-sed}a presents a comparison of the SED between the observed spectra and the best model across a wide wavelength range.
The inset demonstrates that the model provides a remarkably good fit to the $0.4-0.8$\,{\micron} range observed with KOOLS-IFU.

Figure\,\ref{F-sed}b focuses on the \emph{Spitzer}/IRS wavelength range for a detailed comparison.
The model reproduces the observed SED well, except for the $\sim13-20$\,{\micron} range, where an offset is identified. 
The dust formed in O-rich environment is expected to consist of silicates, magnesium/iron oxide grain \citep[e.g., Mg$_{1-x}$Fe$_{x}$O, $x=0-1$;][]{1995A&AS..112..143H}, and aluminum oxide (Al$_{2}$O$_{3}$). 
Al$_{2}$O$_{3}$, with its higher condensation temperature, forms in the inner, high-temperature region of the O-rich circumstellar envelope, 
while silicates condense in regions further out \citep[e.g.,][]{1999A&A...347..594G, 2006A&A...447..553F}. 
Mg$_{1-x}$Fe$_{x}$O forms in the intermediate region between the formation zones of Al$_{2}$O$_{3}$ and silicates. 
Notably, Mg$_{1-x}$Fe$_{x}$O and amorphous Al$_{2}$O$_{3}$ are known to exhibit emission in the $\sim17-18$\,{\micron} range. 
However, its peak position does not coincide with the offset's central wavelength of $\sim16$\,{\micron}. 
This offset is therefore considered to result from the weighting applied to fitting procedure. 
Aside from the $\sim13-20$\,{\micron} range, the SED fitting performs well, and the overall fitting is regarded as successful.

Figures\,\ref{F-sed}c and d show the derived $T_{\rm eff}$, $\log_{10}$\,($L_{\ast}$/L$_{\sun}$), and $\log_{10}\,g$ plotted on 
the theoretical $T_{\rm eff} - \log_{10}$\,($L_{\ast}$/L$_{\sun}$) and $T_{\rm eff} - \log_{10}\,g$ diagrams, respectively.
The derived values lie perfectly on the theoretical post-AGB evolutionary track for 1.0\,M$_{\sun}$ stars with $Z = 0.0028$.
Both \citet{2008ApJ...680.1162H} and \citet{2009ApJ...705..509O} conclude that DdDm\,1 evolved from a $\sim$1\,M$_{\sun}$ star.
Their predicted $T_{\rm eff}/L_{\ast}/\log_{10}\,g$ values are 55000\,K/1000\,L$_{\sun}$/6.5\,cm\,s$^{-2}$ in \citet{2008ApJ...680.1162H}, and 39000\,K/$2000-3000$\,L$_{\sun}$/5.0\,cm\,s$^{-2}$ in \citet{2009ApJ...705..509O}.
It is obvious that neither \citet{2008ApJ...680.1162H} nor \citet{2009ApJ...705..509O} aligns with the predicted post-AGB track for such stars.

The derived $m_{g}$ of 0.287\,M$_{\sun}$ represents the mass within the ionization boundary radius, which includes a small fraction of neutral atomic gas.
In subsection\,\ref{S:d-gasmas}, the progenitor star is expected to have ejected a total mass of $\sim0.35-0.47$\,M$_{\sun}$.
Our model explains $\sim61-82$\,$\%$ of the mass-loss history, with $\sim0.06-0.18$\,M$_{\sun}$ of neutral gas likely lying beyond the ionization front.

How much dust did the progenitor star produce during its evolution?
If the model-predicted GDR of 760 is applicable to the entire nebula including the neutral gas region, and the AGB model predicted total gas mass is $\sim0.35-0.47$\,M$_{\sun}$, 
the progenitor star would have produced $(4.61-6.18)\times10^{-4}$\,M$_{\sun}$ of dust.
By comparison, \citet{2006A&A...447..553F} constructed a theoretical dust production model for stars with an initial mass of 1.0\,M$_{\sun}$ and $Z=0.004$.
Their model predicts that such stars eject a total mass of 0.335\,M$_{\sun}$, including a dust mass of $2.53\times10^{-4}$\,M$_{\sun}$, composed of $1.75\times10^{-4}$\,M$_{\sun}$ silicate 
and $7.80\times10^{-5}$\,M$_{\sun}$ iron dust.
Our derived total dust mass is consistent with \citet{2006A&A...447..553F}, considering the uncertainties in both our {\sc CLOUDY} model and the theoretical model of \citet{2006A&A...447..553F}.

In short, we successfully constructed a highly sophisticated model for DdDm\,1, surpassing previous models in completeness. 
Our model reproduces all observed quantities related to the gas, dust, and central star within a single framework. 
Furthermore, the estimated origin and evolution of the progenitor star of DdDm\,1, as well as the masses of gas and dust ejected during its evolution, are consistent with the latest theoretical models, 
making this achievement particularly noteworthy.

\section{Summary
\label{S-sum}}

We investigated the origin and evolution of the Galactic halo PN DdDm\,1 using the newly obtained Seimei/KOOLS-IFU spectra combined with multiwavelength spectra. 
Through our detailed analyses, we constructed a comprehensive model that elucidates the physical and chemical processes of this PN.

The KOOLS-IFU emission line images achieve $\sim0.9$\,arcsec resolution, resolving the elliptical nebula 
and revealing a compact spatial distribution of the {\feiii} line compared to the {\oiii} line, despite their similar volume emissivities. 
This indicates that iron, with its higher condensation temperature than oxygen, is easily incorporated into dust grains such as silicate, 
making the iron abundance estimate prone to underestimation.
Using a fully data-driven approach, we directly derive ten elemental abundances, the gas-to-dust mass ratio, and the gas and dust masses based on our own distance scale 
and the emitting volumes of gas and dust. 
DdDm\,1 evolved from a star of initial mass of $\sim$1.0\,M$_{\sun}$ in a low-metallicity environment ($Z=0.18$\,Z$_{\sun}$) characterized by a significant carbon deficiency. 
The observed elemental abundances are well consistent with predictions from AGB nucleosynthesis models.
DdDm\,1 is particularly rare due to its low C/O ratio and very low C abundance. 
While other halo PNe with low C/O ratios, such as PRMT\,1 \citep[e.g,][]{1990A&A...237..454P} and NGC\,4361 \citep[e.g.,][]{1990A&A...233..540T}, have been reported, 
DdDm\,1 stands out as one of the most extreme cases, as it evolved into a PN without experiencing efficient TDU during its evolution.
Our photoionization model reproduces all observed quantities in excellent agreement with predictions from AGB nucleosynthesis, post-AGB evolution, and AGB dust production models. 

Our study offers new insights into the evolution and PN formation pathway of low-mass, metal-deficient stars like DdDm\,1 and underscores 
the role of PN progenitors in the chemical enrichment of the Galaxy.

\begin{ack}
I am grateful to the referee for a careful reading and valuable suggestions.

I dedicate this paper to my beloved father, Nobujiro Otsuka. 
Throughout his life, my father stood as my unwavering pillar of strength, believing in me and encouraging me under all circumstances.
He was not only the person I respected the most, but also the foundation of all my achievements and the guiding light that shaped who I am today.
His steadfast love, wisdom, and support enabled me to accomplish all my research.

I also dedicate this paper to my PhD supervisor, Professor Shin'ichi Tamura. 
He was like a parent in my academic journey. 
DdDm\,1 is one of the planetary nebulae he first introduced to me at the start of my graduate studies, 
and it has held a special place in my heart. 
DdDm\,1 is a cherished object that holds the memories of his guidance, wisdom, and the time we shared together.

I was supported by the Japan Society for the Promotion of Science (JSPS) Grants-in-Aids for Scientific Research (C) (19K03914 and 22K03675). 

\end{ack}


\end{document}